\documentclass[11pt]{article}
\usepackage{a4}
\usepackage{times}
\usepackage{amssymb}

\def\beq{\begin{equation}}
\def\bea{\begin{eqnarray}}
\def\eeq{\end{equation}}
\def\eea{\end{eqnarray}}
\def\R{$\mathcal{R}$}
\def\Z{$\mathbb{Z}$}
\def\SM{Standard Model}

\begin{document}
\title{\bf Constructing the supersymmetric Standard Model from intersecting D6-branes on the $\mathbb{Z} _6'$ orientifold }
\maketitle

\vglue 0.35cm
\begin{center}
\author{\bf David Bailin \footnote
{D.Bailin@sussex.ac.uk} \&  Alex Love \\}
\vglue 0.2cm
	{\it  Department of Physics \& Astronomy, University of Sussex\\}
{\it Brighton BN1 9QH, U.K. \\}
\baselineskip=12pt
\end{center}
\vglue 2.5cm
\begin{abstract}
Intersecting stacks of  supersymmetric fractional branes on the \Z$_6'$ orientifold 
may be used to construct the supersymmetric Standard Model.
   If $a,b$  are the   stacks 
  that generate the  $SU(3)_{\rm colour}$ and $SU(2)_L$ gauge particles,  
 then,  in order to obtain {\em just} the chiral spectrum of the (supersymmetric)
  Standard Model (with non-zero Yukawa couplings to the Higgs mutiplets),
   it is necessary that  the number of intersections $a \cap b$ of the stacks $a$ and $b$, and 
  the number of intersections $a \cap b'$ of $a$ with the orientifold image $b'$ of $b$
   satisfy $(a \cap b,a \cap b')=(2,1)$ or $(1,2)$. 
It is also necessary that there is no matter in symmetric representations of the gauge group,
 and not too much matter in antisymmetric representations, on either stack. 
 Fractional branes having  all of these properties may be constructed  on the \Z$_6'$ orientifold. 
We provide a number of new examples having these properties, some of which may be extended to give  the \SM \ spectrum. 
  Specifically, we construct  four-stack
  models with two  further stacks, each 
 with just a single brane, which have the matter spectrum of the  supersymmetric Standard Model, 
 including a single pair of Higgs doublets, {\em plus} three right-chiral neutrino singlets.  Ramond-Ramond tadpole cancellation is achieved by the introduction of background $\bar{H}_3$ flux, the 3-form field strength associated with the Kalb-Ramond 2-form field $B_2$. There remains a single unwanted gauged $U(1)_{B-L}$.
 \end{abstract}

\newpage

\section{Introduction} \label{intro}
An attractive, bottom-up approach to constructing the Standard Model  is to  use intersecting  D6-branes \cite{Lust:2004ks}. 
In these models one starts with 
 two stacks, $a$ and $b$ with $N_a=3$ 
and $N_b=2$, of   D6-branes wrapping the three large spatial 
dimensions plus  3-cycles of the six-dimensional  internal space (typically a torus $T^6$ 
or a Calabi-Yau 3-fold) on which the theory is compactified.
 These generate  the gauge group $U(3) \times U(2) \supset SU(3) _c \times SU(2)_L$, 
 and  the non-abelian component of the standard model gauge group
is immediately assured.
   Further, (four-dimensional) fermions in bifundamental representations 
$({\bf N} _a, \bar{\bf N}_b)= ({\bf 3}, \bar{\bf 2})$ 
of the gauge group can arise at the multiple intersections of the two stacks. 
These are precisely the representations needed for the quark doublets $Q_L$ of the Standard Model, 
and indeed an 
attractive model having just the spectrum of the Standard Model has been constructed \cite{Ibanez:2001nd}. The D6-branes wrap 3-cycles 
of an orientifold $T^6/\Omega$, where $\Omega$ is the world-sheet parity operator. The advantage and, indeed, the necessity of using 
an orientifold stems from the fact that for every stack $a,b, ...$ there is an orientifold image $a',b', ...$. 
At intersections of $a$ and $b$ there are chiral fermions 
in the $({\bf 3}, \bar{\bf 2})$ representation of $U(3) \times U(2)$, where the ${\bf 3}$ has charge $Q_a=+1$ with respect to the 
$U(1)_a$ in $U(3)=SU(3)_{\rm colour} \times U(1)_a$, and the $\bar{\bf 2}$ has charge $Q_b=-1$ with respect to the 
$U(1)_b$ in $U(2)=SU(2)_L \times U(1)_b$.  However, at intersections of $a$ and $b'$ there are chiral fermions 
in the $({\bf 3},{\bf 2})$ representation, where  the ${\bf 2}$ has $U(1)_b$ charge $Q_b=+1$. 
In the model of \cite{Ibanez:2001nd}, the number of intersections $a \cap b$ of the stack $a$ with $b$ is 2, 
and the number of intersections $a \cap b'$ of the stack $a$ with $b'$ is 1. Thus, as required for the \SM , there are 3 quark doublets.
 These have  net $U(1)_a$ charge $Q_a=6$, and net $U(1)_b$ charge $Q_b=-3$. Tadpole cancellation requires that overall both charges,
sum to zero, so further fermions are essential, and indeed required by the \SM. 6 quark-singlet states $u^c_L$ and $d^c_L$ 
belonging to the $({\bf 1}, \bar{\bf 3})$ representation of  $U(1) \times U(3)$, having 
a total of $Q_a=-6$ are sufficient to ensure overall cancellation of $Q_a$, and these arise from the intersections of $a$ with other 
stacks $c,d,...$ having just a single D6-brane. Similarly, 3 lepton doublets $L$, belonging to the $({\bf 2}, \bar{\bf 1})$
 representation of  $U(2) \times U(1)$, having 
a total $U(1)_b$ charge of $Q_b=3$, are sufficient to ensure overall cancellation of $Q_b$, 
and these arise from the intersections of $b$ with other 
stacks having just a single D6-brane. In contrast, had we not used an orientifold, the requirement of 3 quark doublets would 
necessitate having the number of intersections $a \cap b=3$. This makes no difference to the charge $Q_a=6$ carried by the quark doublets,   
but instead the $U(1)_b$ charge carried by the quark doublets is $Q_b=-9$, which cannot be cancelled by just 3 lepton doublets $L$. 
Consequently, additional vector-like fermions are unavoidable unless the orientifold projection is available.
This is why the orientifold is essential if we are to get just the matter content of the \SM \ or of the MSSM. 

 Actually, an orientifold  can allow essentially the standard-model spectrum without vector-like matter even 
when $a \cap b=3$ and $a \cap b'=0$ \cite{Blumenhagen:2001te}. 
This is because in orientifold models it is also possible to get chiral matter in the symmetric and/or antisymmetric representation 
of the relevant gauge group from open strings stretched between a stack and its orientifold image. Both representations have charge $Q=2$ 
with respect to the relevant $U(1)$. The antisymmetric (singlet) representation of $U(2)$ can 
describe a neutrino singlet state $\nu ^c_L$,
 and 3 copies contribute $Q_b=6$ units of $U(1)_b$ charge. If there are also 3 lepton doublets $L$ belonging to the bifundamental representation 
$({\bf 2}, \bar{\bf 1})$
 representation of  $U(2) \times U(1)$, each contributing $Q_b=1$ as above, then the total contribution is $Q_b=9$ which {\bf can} 
 be cancelled by 3 quark doublets $Q_L$ in the $({\bf 3}, \bar{\bf 2})$ representation of $U(3) \times U(2)$. Thus,  
 orientifold models can allow
 the standard-model spectrum plus 3 neutrino singlet states even when $(a \cap b ,a \cap b')=(3,0)$.

Non-supersymmetric intersecting-brane models lead to 
 flavour-changing neutral-current  (FCNC) processes 
that can only be suppressed to levels consistent with the current bounds by making the 
 string scale  rather high, of order $10^4$ TeV, which in turn leads to fine-tuning problems  \cite{Abel:2003yh}.
Further, in  non-supersymmetric theories, such as these, the cancellation of Ramond-Ramond (RR) tadpoles does not ensure 
Neveu Schwarz-Neveu Schwarz (NSNS) tadpole cancellation. 
NSNS tadpoles are simply the first
 derivative of the scalar potential with respect to the scalar fields, specifically the complex structure  and K\"ahler moduli 
 and the dilaton.
  A non-vanishing derivative of the scalar potential signifies that  
 such scalar fields are not even solutions of the equations of motion. 
 Thus a particular consequence of the non-cancellation is that the complex structure moduli are unstable \cite{Blumenhagen:2001mb}. 
 It is well known that the point group of an orbifold fixes the complex structure moduli, so that 
 one way to stabilise these moduli 
  is for the 
D-branes to wrap an orbifold $T^6/P$ rather than a torus $T^6$. 
The FCNC problem can be solved and the complex structure moduli stabilised when the theory is supersymmetric. 
First, a supersymmetric theory is not obliged to have the low string scale that led to problematic FCNCs  induced by string instantons. 
Second, 
 in a supersymmetric theory, RR tadpole cancellation ensures cancellation 
of the NSNS tadpoles \cite{Cvetic:2001tj,Cvetic:2001nr}.
An orientifold is then constructed by quotienting the orbifold with the world-sheet parity operator $\Omega$.
(An orientifold, rather than an orbifold, is required because orientifold O6-planes are needed to allow cancellation of the 
RR charge of the D-branes without using anti-D-branes which would themselves break supersymmetry.) 

In this paper we shall be concerned with the  orientifold having  point group $P=$ \Z$_6'$. We showed in 
a previous paper \cite{Bailin:2006zf} that this {\em does} have (fractional) supersymmetric D6-branes $a$
 and $b$ with intersection numbers 
$(a \cap b, a \cap b')=(1,2)$ or $(2,1)$, which might be used to construct the supersymmetric \SM \ having just the requisite
 standard-model matter content, and in \cite{Bailin:2007va} we presented an example  of just such an extension. 
The 6-torus factorises into three 2-tori as $T^6 = T^2_1 \times T^2_2 \times T^2_3$ with $T^2_k \ (k=1,2,3)$ parametrised by the complex coordinate
 $z_k$. The  generator $\theta$ of the 
point group  $P=$\Z$_6'$ acts on the three complex coordinates $z_k$ as
\beq
\theta z_k= e^{2 \pi iv_k} z_k
\eeq
where
\beq
(v_1,v_2,v_3)= \frac{1}{6}(1,2,-3)
\eeq 
This action must be an automorphism of the lattice, and we take $T^2_1$ and $T^2_2$ to be $SU(3)$ root lattices. Thus the complex structure moduli $U_{1,2}$ 
for $T^2_{1,2}$ are fixed to the values $U_1=U_2= e^{i\pi/3}$. However, 
 since $\theta$ acts on $z_3$ 
as a reflection, the lattice for $T^2_3$, and hence its complex structure $U_3$, is arbitrary.
 The embedding \R \ of the world-sheet parity operator $\Omega$ acts on all $z_k$ as complex conjugation
\beq
\mathcal{R} z_k= \bar{z}_k \quad (k=1,2,3)
\eeq
This too must be an automorphism of the lattice, and this requires the lattice for each torus $T^2_k$ to be in one of two orientations,
{\bf A} or {\bf B}, relative to the Re $z_k$-axis. It also fixes the real part  of the complex structure for $T^2_3$, Re $U_3=0$ for {\bf A} 
and Re $U_3=\frac{1}{2}$ for {\bf B}; the imaginary part remains arbitrary.
We noted in  \cite{Bailin:2006zf} that different orientations of the lattices can give rise to different physics. The realisation of the \SM \ 
presented in the erratum to \cite{Bailin:2007va} utilised the {\bf AAA} configuration.  
 In this paper, we shall present a systematic study of the possibility of constructing 
{\em just} the spectrum  of the \SM \ on all orientations of the lattices. However, since starting this work, it has been shown \cite{Gmeiner:2007zz} that there are no three-generation standard models on this lattice that satisfy the tadpole cancellation conditions.

The fractional branes $\kappa$ with which we are concerned have the general form
\beq
\kappa = \frac{1}{2} \left( \Pi _{\kappa}^ {\rm bulk} + \Pi _{\kappa}^ {\rm ex} \right)
\eeq
where 
\beq
\Pi _{\kappa}^ {\rm bulk}= \sum_{p=1,3,4,6} A^{\kappa}_p \rho _p
\eeq
is an (untwisted) invariant 3-cycle, and
\beq
\Pi _{\kappa}^ {\rm ex}= \sum_{j=1,4,5,6} (\alpha^{\kappa}_j \epsilon _j + \tilde{\alpha}^{\kappa}_j \tilde{\epsilon} _j )
\eeq
is an exceptional 3-cycle associated with the $\theta ^3$-twisted sector. It consists of a collapsed 2-cycle at a 
$\theta^3$ fixed point in $T^2_1 \times T^2_3$ times a 1-cycle in the ($\theta^3$-invariant plane) $T^2_2$. 
The four basis invariant 3-cycles $\rho _p, \ (p=1,3,4,6)$ and the 8 basis exceptional cycles $\epsilon _j$ and 
$\tilde{\epsilon}_j, (j=1,4,5,6)$ are defined in reference \cite{Bailin:2006zf}. Their non-zero intersection numbers are
\bea
\rho _1 \cap \rho _4=4, &\quad & \rho _1 \cap \rho _6=-2 \\
\rho _3 \cap \rho _4=-2, &\quad & \rho _3 \cap \rho _6=4 
\eea
and 
\beq
\epsilon _j \cap \tilde{\epsilon}_k= -2 \delta _{jk}
\eeq
The ``bulk coefficients'' $A^{\kappa}_p$ are given by
\bea
A^{\kappa}_1&=&(n^{\kappa}_1n^{\kappa}_2+n^{\kappa}_1m^{\kappa}_2+m^{\kappa}_1n^{\kappa}_2)n^{\kappa}_3 \\
A^{\kappa}_3&=&(m^{\kappa}_1m^{\kappa}_2+n^{\kappa}_1m^{\kappa}_2+m^{\kappa}_1n^{\kappa}_2)n^{\kappa}_3 \\
A^{\kappa}_4&=&(n^{\kappa}_1n^{\kappa}_2+n^{\kappa}_1m^{\kappa}_2+m^{\kappa}_1n^{\kappa}_2)m^{\kappa}_3 \\
A^{\kappa}_6&=&(m^{\kappa}_1m^{\kappa}_2+n^{\kappa}_1m^{\kappa}_2+m^{\kappa}_1n^{\kappa}_2)m^{\kappa}_3 
\eea
where $(n^{\kappa}_k,m^{\kappa}_k)$ are the (coprime) wrapping numbers for the basis 1-cycles $(\pi _{2k-1}, \pi _{2k})$ of the torus $T^2_k \ (k=1,2,3)$.
The corresponding formulae for the exceptional part are also given in \cite{Bailin:2006zf}.

In the first instance we need two stacks $a$ and $b$ of such fractional branes, with $N_a=3$ and $N_b=2$, satisfying
\beq
(a \cap b, a \cap b')= ({2,1}) \quad {\rm or} \quad (1,2) \label{abab1}
\eeq
 {\it A priori} the weak hypercharge $Y$ is a general linear combination
\beq
Y= \sum_{\kappa} y_{\kappa} Q_{\kappa} \label{Y}
\eeq
of the $U(1)$ charges $Q_{\kappa}$ associated with the stack $\kappa$. We require that both the 
$({\bf 3}, \bar{\bf 2})$ and the $({\bf 3}, {\bf 2})$ representations that occur respectively at the intersections 
of $a$ with $b$ and  with $b'$ have the correct weak hypercharge $Y=1/6$ of the quark doublets $Q_L$. It follows that
\bea
y_a&=& \frac{1}{6} \label{ya}\\
y_b&=&0 \label{yb}
\eea
We also require that both stacks are supersymmetric, which is ensured by two linear conditions
$X^{a,b}>0$ and $Y^{a,b}=0$ on the bulk coefficients $A^{a,b}_p$ for each stack. The precise form of $X^{\kappa}$ and $Y^{\kappa}$ 
depends on the lattice used and is given for all eight possibilities in 
Table 8 of reference \cite{Bailin:2006zf}. In all cases, both $X^{\kappa}$ and $Y^{\kappa}$ depend upon Im $U_3$, so that the requirement of 
supersymmetry on these two stacks, as well as the others that we must add, fixes Im $U_3$. Supersymmetry also 
requires that the exceptional part $\Pi _{\kappa}^ {\rm ex}$ of the stack $\kappa$ is associated with fixed points 
in $T^2_1$ and $T^2_3$ that are traversed by the bulk 3-cycle $\Pi _{\kappa}^ {\rm bulk}$. 
As detailed in \cite{Bailin:2006zf}, the effect of this is 
that, up to Wilson lines, $\Pi _{\kappa}^ {\rm ex}$ is entirely determined by the wrapping numbers 
$(n^{\kappa}_2, m^{\kappa}_2)$ of $\Pi _{\kappa}^ {\rm bulk}$ in $T^2_2$.

In general, besides the gauge supermultiplets that live on each stack $\kappa$, there is also chiral matter in the symmetric ${\bf S}_{\kappa}$ 
and, if $N_{\kappa}>1$, antisymmetric ${\bf A}_{\kappa}$ representations of the gauge group $SU(N_{\kappa})$. For the $a$ stack we have that
${\bf S}_{a}={\bf 6} \in SU(3)_{\rm colour}$, and for the  $b$ stack 
${\bf S}_{b}={\bf 3} \in SU(2)_L$. Both representations are unobserved. Thus
we further require that they do not occur. 
Orientifolding induces  topological defects,  O6-planes, which are sources of RR charge.
The numbers  $\#({\bf S}_{\kappa})$ of symmetric representations and  
$\#({\bf A}_{\kappa})$ of antisymmetric representations are given by
\bea
\#({\bf S}_{\kappa})&=& \frac{1}{2}(\kappa \cap \kappa ' - \kappa \cap \Pi _{\rm O6}) \\
\#({\bf A}_{\kappa})&=& \frac{1}{2}(\kappa \cap \kappa ' + \kappa \cap \Pi _{\rm O6}) 
\eea
where $\Pi _{\rm O6}$ is the homology class of the O6-planes. 
(The required homology classes for all eight lattices are listed in Table 5 of \cite{Bailin:2006zf}.) 
Consequently, the absence of symmetric representations on $a$ and $b$ requires that
\bea
a \cap a'=a \cap \Pi _{\rm O6} \label{noSa}\\
b \cap b'=b \cap \Pi _{\rm O6} \label{noSb}
\eea
For the $a$ stack the antisymmetric representation is 
${\bf A}_{a}=\bar{\bf 3} \in SU(3)_{\rm colour}$ with $Q_a=2$ and hence $Y=1/3$, so that these  representations will be $d$-quark singlet states  $d^c_L$. 
For the  $b$ stack ${\bf A}_{b}={\bf 1} \in SU(2)_L$ with $Q_b=2$ and hence $Y=0$, so any such states will be 
 neutrino  singlets $\nu^c_L$. Clearly, if we are to obtain {\em just} the standard-model spectrum, 
we must not have more than 3 copies of either representation. Hence we must also demand that
\bea 
0 \leq \#({\bf A}_a)= a \cap a' \leq 3 \label{Aa3}  \label{noAa} \\
|\#({\bf A}_b| = |b \cap b'|\leq 3 \label{Ab3}
\eea
As shown in Table 10 of \cite{Bailin:2006zf}, for the lattices in which $T^2_3$ is of {\bf B}-type 
the constraints (\ref{noSa}) and (\ref{noSb}) restrict the wrapping numbers $(n^{a,b}_k, m^{a,b}_k)\bmod 2$ 
for $a$ and $b$ of the basis 1-cycles 
$(\pi _{2k-1}, \pi _{2k})$ on $T^2_k$  for $a$ and $b$ to be in one of {\em two} classes, 
whereas for lattices in which $T^2_3$ is of {\bf A}-type  they must be in 
one of {\em three} classes. This makes the search for solutions satisfying (\ref{abab1}) much easier in the former case than than in the latter. 
It was for this reason that only the former case was considered in \cite{Bailin:2006zf}. In the next section we will present solutions 
satisfying all of the constraints in the cases that $T^2_3$ is of {\bf A}-type. 

As noted earlier, in order to obtain all of the standard-model spectrum, it is necessary to add further stacks $c,d,...$ all consisting of a 
single D6-brane $N_{c,d,...}=1$, so that the gauge group acquires no further non-abelian components. 
 The identification of these additional 
stacks is the main task of this paper. Unlike the (non-abelian)  stacks $a$ and $b$,
there is no requirement that the symmetric representations ${\bf S}_ {c,d,...}$ on these $U(1)$ stacks are absent. (There is no antisymmetric representation of $U(1)$.) 
Such representations are singlets with respect to both of the non-abelian components $SU(3)_{\rm colour}$ and $SU(2)_L$ of the standard-model gauge group 
and might therefore describe  lepton  $\ell ^c_L$ or neutrino $\nu^c_L$ singlet states. 
\section{Quark doublets when $T^2_3$ is of {\bf A}-type}
The objective is to find the (coprime) wrapping numbers $(n^{a,b}_k, m^{a,b}_k)$ for  the two  supersymmetric stacks  $a$ and $b$ of fractional D6-branes that satisfy (\ref{abab1}), (\ref{noSa}), (\ref{noSb}), (\ref{Aa3}) and (\ref{Ab3}). The intersection numbers are given by 
\bea
a \cap b&=& \frac{1}{4} f_{AB}+ \frac{1}{4}(i^a_1,i^a_2)(j^a_1,j^a_2) \cap (i^b_1,i^b_2)(j^b_1,j^b_2) \\
a \cap b'&=& \frac{1}{4} f_{AB'}+ \frac{1}{4}(i^a_1,i^a_2)(j^a_1,j^a_2) \cap (i^b_1,i^b_2)(j^b_1,j^b_2)'
\eea
where we are using the notation for the exceptional parts used previously
\beq
\Pi _{a(i^a_1,i^a_2)(j^a_1,j^a_2)}^{\rm ex}(n^a_2,m^a_2) \rightarrow (i^a_1,i^a_2)(j^a_1,j^a_2)
\eeq
The contributions from the bulk parts are
\bea
f_{AB}& \equiv &\Pi _a^{\rm bulk} \cap \Pi _b^{\rm bulk} \\
&=& 4(A^a_1A^b_4-A^a_4A^b_1)-2(A^a_1A^b_6-A^a_6A^b_1)-2(A^a_3A^b_4-A^a_4A^b_3)+4(A^a_3A^b_6-A^a_6A^b_3) \\
f_{AB'}& \equiv &\Pi _a^{\rm bulk} \cap {\Pi _b^{\rm bulk} }'
\eea
and the function $-f_{AB'}$  is given in Table 11 of \cite{Bailin:2006zf} for the various lattices. 
(The sign change from the Table is a consequence of the overall sign change for intersections of the bulk 3-cycles, 
as explained in the Erratum.)

As in \cite{Bailin:2006zf}, by acting with the generator $\theta$ of the point group $\mathbb{Z}_6'$ on the wrapping numbers $(n^{a,b}_1, m^{a,b}_1)$ on $T^2_1$, we may take $(n^{a}_1, m^{a}_1)=(n^{a}_3, m^{a}_3) \bmod 2$, and likewise for $b$. Since there are  three possibilities  for $(n^{a}_1, m^{a}_1) \bmod 2$, namely $(1,0), \ (0,1), \ {\rm or} \ (1,1) \bmod 2$ when $(n^{a}_1, m^{a}_1)$ are coprime, there are nine distinct pairs for $(n^{a}_1, m^{a}_1)(n^{b}_1, m^{b}_1)\bmod 2$, three with $(n^{a}_1, m^{a}_1)=(n^{b}_1, m^{b}_1) \bmod 2$, and six with 
$(n^{a}_1, m^{a}_1)\neq (n^{b}_1, m^{b}_1) \bmod2 $.  When $T^2_3$ is of {\bf B}-type, we showed that we need only consider the  cases in which $(n^{a}_1, m^{a}_1)=(n^{a}_3, m^{a}_3)=(1,0), \ {\rm or} \ (1,1) \bmod 2$, and the calculation of the contribution $(i^a_1,i^a_2)(j^a_1,j^a_2) \cap (i^b_1,i^b_2)(j^b_1,j^b_2)$ of the exceptional branes to $a \cap b$ for these cases is presented in \S 6 of \cite{Bailin:2006zf}; the  calculation of  the  corresponding contributions to $a \cap b'$ for the four lattices in which $T^2_3$ is of {\bf B}-type is given in the appendices of that paper\footnote{
Again, as explained in the Erratum, there is an overall sign change for all calculations of $(i^a_1,i^a_2)(j^a_1,j^a_2) \cap (i^b_1,i^b_2)(j^b_1,j^b_2)'$ presented in the Appendices.}. 
 To deal with the cases in which $T^2_3$ is of {\bf A}-type,  we  therefore need only present the  contributions from the exceptional branes to $a \cap b$  when $(n^{a}_1, m^{a}_1)=(n^{a}_3, m^{a}_3)=(0,1) \bmod 2$ and/or $(n^{b}_1, m^{b}_1)=(n^{b}_3, m^{b}_3)=(0,1) \bmod 2$; the contributions from the exceptional branes to $a \cap b'$ for the four lattices in which $T^2_3$ is of {\bf A}-type are given in the appendices.
\subsection{$(n^{a,b}_1,m^{a,b}_1)=(n^{a,b}_3,m^{a,b}_3)=(0,1)  \bmod 2$} \label{01}
In this case $(i^a_1,i^a_2),(i^b_1,i^b_2)=(46)$ and $(j^a_1,j^a_2),(j^b_1,j^b_2)=(15)$ or $(46)$.
\bea
(46)(15) &\cap &  (46)(15) =(46)(46) \cap (46)(46) =\nonumber \\
&=& (-1)^{\tau ^a _0 +\tau ^b _0 +1}2[1+(-1)^{\tau ^a_2 +\tau ^b_2}]\left[ (m^a_2n^b_2-n^a_2m^b_2)[1+ (-1)^{\tau ^a _1 +\tau ^b_1}] \right.+ \nonumber \\
&+& \left.(-1)^{\tau ^a_1}(n^a_2n^b_2+m^a_2m^b_2+n^a_2m^b_2)+(-1)^{\tau ^b_1+1}(n^a_2n^b_2+m^a_2m^b_2+m^a_2n^b_2) \right] \\
(46)(15) &\cap &  (46)(46) =0 
\eea
\subsection{$(n^{a}_1, m^{a}_1)=(n^{a}_3, m^{a}_3)=(1,1) \bmod 2$, \ $(n^{b}_1, m^{b}_1)=(n^{b}_3, m^{b}_3)=(0,1) \bmod 2$} \label{1101}
In this case $(i^a_1,i^a_2)=(46), \ (j^a_1,j^a_2)=(15)$ or $(46)$, and  $(i^b_1,i^b_2)=(45), \ (j^b_1,j^b_2)=(16)$ or $(45)$.
\bea
(45)(16) &\cap &  (46)(15) =(-1)^{\tau ^a_2+ \tau ^b_2}(45)(16)\cap (46)(46)= \nonumber \\
&=& (45)(45) \cap (46)(46)=(-1)^{\tau ^a_2+ \tau ^b_2}(45)(45) \cap (46)(15) =\nonumber \\
&=& (-1)^{\tau ^a _0 +\tau ^b _0 +1}2 \left[(m^a_2n^b_2-n^a_2m^b_2) 
- [(-1)^{\tau ^a_1}+(-1)^{\tau ^b_1}](n^a_2n^b_2+m^a_2m^b_2+m^a_2n^b_2)+\right. \nonumber \\
&+&\left.(-1)^{\tau ^a_1+\tau ^b_1}(n^a_2n^b_2+m^a_2m^b_2+n^a_2m^b_2)           \right]  
\eea
Interchanging the labels $ a \leftrightarrow b$  in this calculation immediately gives the results for the case when $(n^{a}_1, m^{a}_1)=(n^{a}_3, m^{a}_3)=(0,1) \bmod 2$, \ $(n^{b}_1, m^{b}_1)=(n^{b}_3, m^{b}_3)=(1,1) \bmod 2$.
\subsection{$(n^{a}_1, m^{a}_1)=(n^{a}_3, m^{a}_3)=(0,1) \bmod 2$, \ $(n^{b}_1, m^{b}_1)=(n^{b}_3, m^{b}_3)=(1,0) \bmod 2$} \label{0110}
In this case $(i^a_1,i^a_2)=(46), \ (j^a_1,j^a_2)=(15)$ or $(46)$, and  $(i^b_1,i^b_2)=(45), \ (j^b_1,j^b_2)=(16)$ or $(45)$.
\bea
(46)(15) &\cap &  (56)(14) =(-1)^{\tau ^a_2}(46)(15)\cap (56)(56)= \nonumber \\
&=& (-1)^{\tau ^a_2+ \tau ^b_2}(46)(46) \cap (56)(56)=(-1)^{\tau ^b_2}(46)(46) \cap (56)(14) =\nonumber \\
&=& (-1)^{\tau ^a _0 +\tau ^b _0 +1}2 \left[(n^a_2n^b_2+m^a_2m^b_2+n^a_2m^b_2) \right. + \nonumber \\
&+&\left. [(-1)^{\tau ^a_1+1}+(-1)^{\tau ^b_1+1}](n^a_2n^b_2+m^a_2m^b_2+m^a_2n^b_2)+
(-1)^{\tau ^a_1+\tau ^b_1}(m^a_2n^b_2-n^a_2m^b_2)           \right]  
\eea
As above, interchanging the labels $ a \leftrightarrow b$  in this calculation immediately gives the results for the case when $(n^{a}_1, m^{a}_1)=(n^{a}_3, m^{a}_3)=(1,0) \bmod 2$, \ $(n^{b}_1, m^{b}_1)=(n^{b}_3, m^{b}_3)=(0,1) \bmod 2$.
\section{Computations when $T^2_3$ is of {\bf A}-type} \label{T32a}
Using the calculations presented in the previous section and the appendices, we seek wrapping numbers 
$(n_k^{a,b},m_k^{a,b}) \ (k=1,2,3)$ for two stacks $a$ and $b$ of fractional branes that yield the required intersection numbers $(a \cap b,a \cap b')=(1,2) \ {\rm or} \ (2,1)$,
 that have no symmetric matter on $a$ or $b$, that satisfy the supersymmetry
 constraints $Y^a, \ Y^b=0$ and $X^{a},\  X^b>0$, 
 and that do not have more than three copies of  matter in the antisymmetric representation on $a$ or $b$.
\subsection{AAA lattice} \label{aaal}
On the {\bf AAA} lattice the supersymmetry constraints for a general stack $\kappa$ are
\bea
X^{\kappa} &\equiv& 2A^{\kappa}_1-A^{\kappa} _3-A^{\kappa}_6\sqrt{3} \ {\rm Im} \ U_3>0 \label{Xaaa}\\
Y^a &\equiv& \sqrt{3} A^a_3+(2A^a_4-A^a_6) {\rm Im} \ U_3=0 \label{Yaaa}
\eea
 We found solutions with the required properties for four values of 
\bea
{\rm Im} \ U_3 &=& -\frac{1}{ \sqrt{3}} \label{imu31} \\
&=& -{ \sqrt{3}} \label{imu32} \\
&=& -\frac{2}{ \sqrt{3}} \label{imu33} \\
&=& -\frac{1}{2 \sqrt{3}} \label{imu34} 
\eea
These are displayed in  Tables \ref{aaa1}, \ref{aaa2}, \ref{aaa3} and \ref{aaa4} respectively.
\begin{table}
 \begin{center}
\begin{tabular}{||c|c|c||c|c|c||} \hline \hline
$(n^a_1,m^a_1;n^a_2,m^a_2;n^a_3,m^a_3)$&$(A^a_1,A^a_3,A^a_4,A^a_6)$&$\#({\bf A}_a)$&$(n^b_1,m^b_1;n^b_2,m^b_2;n^b_3,m^b_3)$&$(A^b_1,A^b_3,A^b_4,A^b_6)$&$\#({\bf A}_b)$  \\ \hline \hline
$(1,-1;1,0;1,-3)$&$(0,-1,0,3)$&0&$(1,0;1,0;1,0)$&$(1,0,0,0)$&0 \\ \hline
$(1,-1;1,1;1,-1)$&$(1,-1,-1,1)$&0&$(1,0;1,0;1,0)$&$(1,0,0,0)$&0 \\ \hline
$(1,1;1,-1;1,-1)$&$(1,-1,-1,1)$&0 &$(1,0;1,0;1,0)$&$(1,0,0,0)$&0 \\ \hline \hline
$(1,-1;-1,1;1,3)$&$(1,1,3,3)$&0&$(-2,1;1,-1;0,1)$&$(0,0,1,2)$&0 \\ \hline


$(1,-1;-1,2;1,1)$&$(2,1,2,1)$&0&$(-2,1;1,-1;0,1)$&$(0,0,1,2)$&0 \\ \hline
$(1,1;1,0;1,1)$&$(2,1,2,1)$&0&$(-2,1;1,-1;0,1)$&$(0,0,1,2)$&0 \\ \hline \hline

 \end{tabular}
\end{center} 
\caption{ \label{aaa1} Solutions on the {\bf AAA} lattice with ${\rm Im} \ U_3=-1/\sqrt{3}$.}
 \end{table} 
\begin{table}
 \begin{center}
\begin{tabular}{||c|c|c||c|c|c||} \hline \hline
$(n^a_1,m^a_1;n^a_2,m^a_2;n^a_3,m^a_3)$&$(A^a_1,A^a_3,A^a_4,A^a_6)$&$\#({\bf A}_a)$&$(n^b_1,m^b_1;n^b_2,m^b_2;n^b_3,m^b_3)$&$(A^b_1,A^b_3,A^b_4,A^b_6)$&$\#({\bf A}_b)$  \\ \hline \hline
$(1,-1;-1,1;1,1)$&$(1,1,1,1)$&2&$(-2,1;1,-1;0,1)$&$(0,0,1,2)$&0 \\ \hline \hline
\end{tabular}
\end{center} 
\caption{ \label{aaa2} Solution on the {\bf AAA} lattice with ${\rm Im} \ U_3=-\sqrt{3}$.}
 \end{table}  
\begin{table}
 \begin{center}
\begin{tabular}{||c|c|c||c|c|c||} \hline \hline
$(n^a_1,m^a_1;n^a_2,m^a_2;n^a_3,m^a_3)$&$(A^a_1,A^a_3,A^a_4,A^a_6)$&$\#({\bf A}_a)$&$(n^b_1,m^b_1;n^b_2,m^b_2;n^b_3,m^b_3)$&$(A^b_1,A^b_3,A^b_4,A^b_6)$&$\#({\bf A}_b)$  \\ \hline \hline
$(0,1;0,-1;2,-3)$&$(0,-2,0,3)$&-3&$(1,0;-1,0;-1,0)$&$(1,0,0,0)$&0 \\ \hline
\end{tabular}
\end{center} 
\caption{ \label{aaa3} Solution on the {\bf AAA} lattice with ${\rm Im} \ U_3=-2/\sqrt{3}$.}
 \end{table} 
\begin{table}
 \begin{center}
\begin{tabular}{||c|c|c||c|c|c||} \hline \hline
$(n^a_1,m^a_1;n^a_2,m^a_2;n^a_3,m^a_3)$&$(A^a_1,A^a_3,A^a_4,A^a_6)$&$\#({\bf A}_a)$&$(n^b_1,m^b_1;n^b_2,m^b_2;n^b_3,m^b_3)$&$(A^b_1,A^b_3,A^b_4,A^b_6)$&$\#({\bf A}_b)$  \\ \hline \hline
$(1,-2;1,-1;-1,-2)$&$(2,1,4,2)$&-3&$(-2,1;1,-1;0,1)$&$(0,0,1,2)$&0\\ \hline
$(1,0;1,1;1,2)$&$(2,1,4,2)$&-3&$(-2,1;1,-1;0,1)$&$(0,0,1,2)$&0 \\ \hline \hline
\end{tabular}
\end{center} 
\caption{ \label{aaa4} Solutions on the {\bf AAA} lattice with ${\rm Im} \ U_3=-1/2\sqrt{3}$.}
 \end{table}

On this lattice and on the others in which $T^2_3$ is of {\bf A}-type, and indeed on the lattices in which $T^2_3$ is of {\bf B}-type,
 it appears that solutions only arise when 
$(n^{a}_1, m^{a}_1)=(n^{a}_3, m^{a}_3)\bmod 2 \neq (n^{b}_1, m^{b}_1)=(n^{b}_3, m^{b}_3) \bmod 2$. 

\subsubsection{Solutions with ${\rm Im} \ U_3=-1/\sqrt{3}$}
The first solution in Table \ref{aaa1} has  $SU(3)_{\rm colour}$ stack $a$, with  
 bulk part given by
\bea
{\Pi_a^{\rm bulk}}=&=&  -\rho_3 +3 \rho _6  \label{abulkaaa1} \\
{\Pi_a^{\rm bulk}}'&=&\rho _1+\rho_3+3 \rho _4 +3 \rho _6 \label{a1bulkaaa1}
\eea
From Table 5 of \cite{Bailin:2006zf} the ${\rm O6}$-plane is 
\beq
\Pi _{\rm O6}= \rho_1+\rho _4 +2\rho_6
\eeq
on the  {\bf AAA} lattice. 
Hence, 
\bea
 \Pi_a^{\rm bulk}\cap \Pi _{\rm O6}&=&0  \\
\Pi_a^{\rm bulk} \cap {\Pi_a^{\rm bulk}}'&=&-12
\eea 
Since $(n^a_2,m^a_2)=(1,0)$, the exceptional part of $a$ is
\bea
\Pi_a^{\rm ex}&=&(45)(16) (n^a_2,m^a_2) \nonumber \\
&=&(-1)^{\tau ^a_0}\left([-(-1)^{\tau _1^a}] [ \epsilon _1+ (-1)^{\tau ^a_2}\epsilon_6] 
+[1-(-1)^{\tau _1^a}][\tilde {\epsilon} _1+ (-1)^{\tau ^a_2}\tilde{\epsilon}_6] \right)    \label{aex16}
\eea
In both cases 
\beq
 \Pi_a^{\rm ex}\cap {\Pi_a^{\rm ex}}' =4[1-2(-1)^{\tau ^a_1}]
\eeq
and the absence of symmetric representations  on $a$ is guaranteed provided that
\beq
\tau ^a_1=1 \bmod 2 \label{taua1}
\eeq
Hence, 
\bea
\Pi_a^{\rm ex}&=&(-1)^{\tau ^a_0 }\left([\epsilon _1+(-1)^{\tau ^a_2}\epsilon _6] 
+2[\tilde{\epsilon}_1+(-1)^{\tau ^a_2}\tilde{\epsilon}_6]  \right)
\label{piaex1} 
\eea

The $SU(2)_L$ stack $b$ has
\beq
\Pi_b^{\rm bulk}=\rho _1 = {\Pi_b^{\rm bulk}}' \label{pib1}
\eeq
 Hence
\bea
 \Pi_b^{\rm bulk}\cap \Pi _{\rm O6}&=&0  \\
\Pi_b^{\rm bulk} \cap {\Pi_b^{\rm bulk}}'&=&0
\eea 
Since $(n^b_2,m^b_2)=(1,0)$, the exceptional part is given by
\bea
\Pi_b^{\rm ex}&=&(56)(14)(n^b_2,m^b_2)=(-1)^{\tau ^b_0}\left( [(-1)^{\tau ^b_1}-1][ \epsilon _1+ (-1)^{\tau ^b_2}\epsilon_4] 
-[\tilde {\epsilon} _1+ (-1)^{\tau ^b_2}\tilde{\epsilon}_4] \right)    \label{bex14}
\eea
The orientifold
image is given by
\beq
{\Pi_b^{\rm ex}}'=(-1)^{\tau ^b_1+1}\Pi_b^{\rm ex}
\eeq
Hence, 
\beq
\Pi_b^{\rm ex} \cap {\Pi_b^{\rm ex}}'=0
\eeq
and the absence of symmetric representations on $b$ is guaranteed independently of the  choice of $\tau ^b_1$.

The contributions to $a \cap b$ and $a \cap b'$ from the bulk parts are 
\beq
(\Pi _a^{\rm bulk} \cap \Pi _b^{\rm bulk}, \Pi _a^{\rm bulk} \cap {\Pi _b^{\rm bulk}}')=(6,6) \label{abbulk}
\eeq
so that the required
 intersection numbers $(a \cap b, a \cap b') =(1,2)$ or $(2,1)$ are  achieved when 
\beq
 (\Pi _a^{\rm ex} \cap \Pi _b^{\rm ex}, \Pi _a^{\rm ex} \cap {\Pi _b^{\rm ex}}')= \pm (2,-2) \label{aexbex}
 \eeq
 From  (\ref{aex16}) and  (\ref{bex14}) with (\ref{taua1})   we find that
 \beq
\Pi _a^{\rm ex} \cap \Pi _b^{\rm ex}=(-1)^{\tau ^a_0 + \tau ^b_0}2[2(-1)^{\tau _1^b}-1]= (-1)^{\tau _1^b+1} \Pi _a^{\rm ex} \cap {\Pi _b^{\rm ex}}'
\eeq  
Thus (\ref{aexbex}) requires that
\beq
\tau ^b_1=0 \label{taub0}
\eeq
Thus in this solution the $SU(2)_L$ stack   $b$ has  
\bea
\Pi_b^{\rm ex}&=& -{\Pi_b^{\rm ex}}' \\
&=&(-1)^{\tau ^b_0+1} [ \tilde{\epsilon} _1+ (-1)^{\tau ^b_2}\tilde{\epsilon}_4]  \label{pibex} 
\eea

The second and third solutions have the same $SU(2)_L$ stack $b$ as in the first solution,
 but different $SU(3)_{\rm colour}$ stacks $a$. For the second solution, proceeding similarly, 
we find
\bea
\Pi_a^{\rm ex}&=&(-1)^{\tau ^a_0 }\left(-[\epsilon _1+(-1)^{\tau ^a_2}\epsilon _6] 
+[\tilde{\epsilon}_1+(-1)^{\tau ^a_2}\tilde{\epsilon}_6]  \right)
\label{piaex2} 
\eea
and for the third
\bea
\Pi_a^{\rm ex}&=&(-1)^{\tau ^a_0+1 }\left([\epsilon _1+(-1)^{\tau ^a_2}\epsilon _6] 
+[\tilde{\epsilon}_1+(-1)^{\tau ^a_2}\tilde{\epsilon}_6]  \right)
\label{piaex3} 
\eea

 The three solutions displayed in the lower half of Table \ref{aaa1} have $SU(3)_{\rm colour}$ stacks $a$ that (up to a phase) are 
the orientifold duals of the  solutions in the upper half of the Table. We get
\bea
\Pi_a^{\rm ex}&=&(-1)^{\tau ^a_0 }\left([\epsilon _1+(-1)^{\tau ^a_2}\epsilon _6] 
-[\tilde{\epsilon}_1+(-1)^{\tau ^a_2}\tilde{\epsilon}_6]  \right) \label{aex4aaa}\\
&=&(-1)^{\tau ^a_0 }\left([\epsilon _1+(-1)^{\tau ^a_2}\epsilon _6] 
+2[\tilde{\epsilon}_1+(-1)^{\tau ^a_2}\tilde{\epsilon}_6]  \right) \\
\label{piaex12} 
&=&(-1)^{\tau ^a_0 +1}[\epsilon _1+(-1)^{\tau ^a_2}\epsilon _6] 
\eea
respectively. They have the same $SU(2)_L$ stack $b$ with 
\bea
\Pi_b^{\rm ex}&=& -{\Pi_b^{\rm ex}}' \label{pibx1}\\
&=&(-1)^{\tau ^b_0} [ \tilde{\epsilon} _1+ (-1)^{\tau ^b_2}\tilde{\epsilon}_5]  \label{pibex456} 
    \eea
\subsubsection{{Solution with ${\rm Im} \ U_3=-\sqrt{3}$}} \label{saaa2}
The absence of symmetric representations on the $SU(3)_{\rm colour}$ stack $a$ for the solution given in Table \ref{aaa2} 
requires that
\beq
\tau ^a_1=0 \bmod 2
\eeq
Then
\bea
\Pi_a^{\rm ex}&=&(45)(16)(n^a_2,m^a_2)=(-1)^{\tau ^a_0}\left( [ \epsilon _1+ (-1)^{\tau ^a_2}\epsilon_6] 
+[\tilde {\epsilon} _1+ (-1)^{\tau ^a_2}\tilde{\epsilon}_6] \right)    \label{aex1621}
\eea
The $SU(2)_L$ stack $b$ is identical to that given in (\ref{pibex456}) for the three solutions 
 in the bottom half of Table \ref{aaa1}.
\subsubsection{Solution with ${\rm Im} \ U_3=-2/\sqrt{3}$}
The absence of symmetric representations on the $SU(3)_{\rm colour}$ stack $a$ for the solution given in Table \ref{aaa3} 
requires that
\beq
\tau ^a_1=1 \bmod 2
\eeq
Then
\bea
\Pi_a^{\rm ex}&=&(-1)^{\tau ^a_0}\left( [ \epsilon _1+ (-1)^{\tau ^a_2}\epsilon_5] 
+2[\tilde {\epsilon} _1+ (-1)^{\tau ^a_2}\tilde{\epsilon}_5] \right) 
\eea
The $SU(2)_L$ stack $b$ is identical to that given in (\ref{pibex}) for the three solutions 
 in the top half of Table \ref{aaa1}.
\subsubsection{Solution with ${\rm Im} \ U_3=-1/2\sqrt{3}$}
The absence of symmetric representations on the $SU(3)_{\rm colour}$ stack $a$ for the  first solution given in Table \ref{aaa4} 
requires that
\beq
\tau ^a_1=1 \bmod 2
\eeq
Then
\bea
\Pi_a^{\rm ex}&=&(-1)^{\tau ^a_0+1}\left( [ \epsilon _1+ (-1)^{\tau ^a_2}\epsilon_4] +
2[\tilde {\epsilon} _1+ (-1)^{\tau ^a_2}\tilde{\epsilon}_4] \right) \label{aexaaa42}
\eea
The $SU(2)_L$ stack $b$ is identical to that given in (\ref{pibex456}) for the three solutions 
 in the bottom half of Table \ref{aaa1}.

The second solution in Table \ref{aaa4} differs from the first only in the wrapping numbers $(n^a_2,m^a_2)$ of the 
$SU(3)_{\rm colour}$ stack $a$. The absence of symmetric representastions on $a$ then requires that $\tau ^a_1=0 \bmod 2$ 
for this solution, but then $\Pi _a ^{\rm ex}$ is identical to that given in (\ref{aexaaa42}). Thus this solution is identical to the first.

\subsection{BAA lattice} \label{baal}
On the {\bf BAA} lattice the supersymmetry constraints for a general stack $\kappa$ are
\bea
X^{\kappa} &\equiv& \sqrt{3}A^{\kappa}_1+(A^{\kappa} _4-2A^{\kappa}_6 )\ {\rm Im} \ U_3>0  \label{Xbaa}\\
Y^a &\equiv&  2A^{\kappa}_3-A^{\kappa}_1+A^{\kappa}_4\sqrt{3}\ {\rm Im} \ U_3=0 \label{Ybaa}
\eea
 We again found solutions with the required properties for four values of 
\bea
{\rm Im} \ U_3 &=& -\frac{1}{ \sqrt{3}} \label{imu31baa} \\
&=& -{ \sqrt{3}} \label{imu32baa} \\
&=& -{2}{ \sqrt{3}} \label{imu33baa} \\
&=& -\frac{\sqrt{3}}{2 } \label{imu34baa} 
\eea
These are displayed in  Tables \ref{baa1}, \ref{baa2}, \ref{baa3} and \ref{baa4} respectively.
\begin{table}
 \begin{center}
\begin{tabular}{||c|c|c||c|c|c||} \hline \hline
$(n^a_1,m^a_1;n^a_2,m^a_2;n^a_3,m^a_3)$&$(A^a_1,A^a_3,A^a_4,A^a_6)$&$\#({\bf A}_a)$&$(n^b_1,m^b_1;n^b_2,m^b_2;n^b_3,m^b_3)$&$(A^b_1,A^b_3,A^b_4,A^b_6)$&$\#({\bf A}_b)$  \\ \hline \hline
$(1,-1;0,1;1,-1)$&$(1,0,-1,0)$&2 &$(1,-2;-1,1;1,0)$&$(2,1,0,0)$&0\\ \hline \hline
 \end{tabular}
\end{center} 
\caption{ \label{baa1} Solution  on the {\bf BAA} lattice with ${\rm Im} \ U_3=-1/\sqrt{3}$.}
 \end{table} 
\begin{table}
 \begin{center}
\begin{tabular}{||c|c|c||c|c|c||} \hline \hline
$(n^a_1,m^a_1;n^a_2,m^a_2;n^a_3,m^a_3)$&$(A^a_1,A^a_3,A^a_4,A^a_6)$&$\#({\bf A}_a)$&$(n^b_1,m^b_1;n^b_2,m^b_2;n^b_3,m^b_3)$&$(A^b_1,A^b_3,A^b_4,A^b_6)$&$\#({\bf A}_b)$  \\ \hline \hline
$(1,-1;-2,1;1,1)$&$(1,2,1,2)$&0&$(0,1;0,1,0,1)$&$(0,0,0,1)$&0 \\ \hline
$(1,-1;-1,1;3,1)$&$(3,3,1,1)$&0&$(0,1;0,1;0,1)$&$(0,0,0,1)$&0 \\ \hline \hline
$(1,-1;1,1;1,-1)$&$(1,-1,-1,1)$&0&$(1,-2;-1,1;1,0)$&$(2,1,0,0)$&0\\ \hline
$(1-1;0,1;3,-1)$&$(3,0,-1,0)$&0&$(1,-2;-1,1;1,0)$&$(2,1,0,0)$&0\\ \hline \hline
\end{tabular}
\end{center} 
\caption{ \label{baa2} Solutions on the {\bf BAA} lattice with ${\rm Im} \ U_3=-\sqrt{3}$.}
 \end{table} 
\begin{table}
 \begin{center}
\begin{tabular}{||c|c|c||c|c|c||} \hline \hline
$(n^a_1,m^a_1;n^a_2,m^a_2;n^a_3,m^a_3)$&$(A^a_1,A^a_3,A^a_4,A^a_6)$&$\#({\bf A}_a)$&$(n^b_1,m^b_1;n^b_2,m^b_2;n^b_3,m^b_3)$&$(A^b_1,A^b_3,A^b_4,A^b_6)$&$\#({\bf A}_b)$  \\ \hline \hline
$(0,1;-1,2;-2,1)$&$(2,-2,-1,1)$&-3&$(1,-2;-1,1;1,0)$&$(2,1,0,0)$&0 \\ \hline \hline
\end{tabular}
\end{center} 
\caption{ \label{baa3} Solution on the {\bf BAA} lattice with ${\rm Im} \ U_3=-2\sqrt{3}$.}
 \end{table} 
\begin{table}
 \begin{center}
\begin{tabular}{||c|c|c||c|c|c||} \hline \hline
$(n^a_1,m^a_1;n^a_2,m^a_2;n^a_3,m^a_3)$&$(A^a_1,A^a_3,A^a_4,A^a_6)$&$\#({\bf A}_a)$&$(n^b_1,m^b_1;n^b_2,m^b_2;n^b_3,m^b_3)$&$(A^b_1,A^b_3,A^b_4,A^b_6)$&$\#({\bf A}_b)$  \\ \hline \hline
$(1,0;0,1;3,2)$&$(3,3,2,2)$&-3&$(0,1;0,1;0,1)$&$(0,0,0,1)$&0 \\ \hline \hline
\end{tabular}
\end{center} 
\caption{ \label{baa4} Solution  on the {\bf BAA} lattice with ${\rm Im} \ U_3=-\sqrt{3}/2$.}
 \end{table}
\subsubsection{Solution with ${\rm Im} \ U_3=-1/\sqrt{3}$} \label{sbaa1}
From Table 5 of \cite{Bailin:2006zf} the ${\rm O6}$-plane is 
\beq
\Pi _{\rm O6}= 2\rho_1+\rho _3 +\rho_6
\eeq
on the  {\bf BAA} lattice. 
Hence, for the solution displayed in Table \ref{baa1}
\bea
 \Pi_a^{\rm bulk}\cap \Pi _{\rm O6}&=&4  \\
\Pi_a^{\rm bulk} \cap {\Pi_a^{\rm bulk}}'&=&4 \\
\Pi_a^{\rm ex}\cap {\Pi_a^{\rm ex}}' &=&-4[1-2(-1)^{\tau ^a_1}]
\eea 
and the absence of symmetric representations  on $a$ is guaranteed provided that
\beq
\tau ^a_1=0 \bmod 2 \label{taua1baa}
\eeq
Hence, 
\bea
\Pi_a^{\rm ex}&=&(-1)^{\tau ^a_0 }[\tilde{\epsilon}_1+(-1)^{\tau ^a_2}\tilde{\epsilon}_6] 
\label{piaex1baa} 
\eea

The $SU(2)_L$ stack $b$ has
\beq
\Pi_b^{\rm bulk}=2\rho _1+\rho_3 = {\Pi_b^{\rm bulk}}' \label{pib2}
\eeq
 Hence
\bea
 \Pi_b^{\rm bulk}\cap \Pi _{\rm O6}&=&0  \\
\Pi_b^{\rm bulk} \cap {\Pi_b^{\rm bulk}}'&=&0
\eea 
Since $(n^b_2,m^b_2)=(-1,1)$, the exceptional part is given by
\bea
\Pi_b^{\rm ex}&=&(56)(14)(n^b_2,m^b_2) \nonumber \\
&=&(-1)^{\tau ^b_0}\left( -(-1)^{\tau ^b_1}[ \epsilon _1+ (-1)^{\tau ^b_2}\epsilon_4] 
+[1-(-1)^{\tau ^b_1}][\tilde {\epsilon} _1+ (-1)^{\tau ^b_2}\tilde{\epsilon}_4] \right)    \label{bex14baa}
\eea
The orientifold
image is given by
\beq
{\Pi_b^{\rm ex}}'=(-1)^{\tau ^b_1+1}\Pi_b^{\rm ex}
\eeq
Hence, 
\beq
\Pi_b^{\rm ex} \cap {\Pi_b^{\rm ex}}'=0
\eeq
and the absence of symmetric representations on $b$ is guaranteed independently of the  choice of $\tau ^b_1$.

The contributions to $a \cap b$ and $a \cap b'$ from the bulk parts again satisfy (\ref{abbulk}) 
so that the required
 intersection numbers $(a \cap b, a \cap b') =(1,2)$ or $(2,1)$ are  achieved when 
(\ref{aexbex}) is satisfied.
 From  (\ref{piaex1baa}) and  (\ref{bex14baa}) with (\ref{taua1baa})   we find that
 \beq
\Pi _a^{\rm ex} \cap \Pi _b^{\rm ex}=(-1)^{\tau ^a_0 + \tau ^b_0}2(-1)^{\tau _1^b}= (-1)^{\tau _1^b+1} \Pi _a^{\rm ex} \cap {\Pi _b^{\rm ex}}'
\eeq  
Thus (\ref{aexbex}) requires that (\ref{taub0}) is satisfied, so that 
\bea
\Pi_b^{\rm ex}&=& -{\Pi_b^{\rm ex}}' \nonumber \\
&=&(-1)^{\tau ^b_0+1} [ {\epsilon} _1+ (-1)^{\tau ^b_2}{\epsilon}_4]  \label{bex141}
\eea
\subsubsection{Solutions with ${\rm Im} \ U_3=-\sqrt{3}$} \label{sbaa2}
The absence of symmetric representations on the $SU(3)_{\rm colour}$ stack $a$ of the 
first solution in Table \ref{baa2} requires that $\tau ^a_1=0 \bmod 2$ so that
\bea
\Pi_a^{\rm ex}&=&(-1)^{\tau ^a_0 }\left( 2[{\epsilon}_1+(-1)^{\tau ^a_2}{\epsilon}_6] +
[\tilde{\epsilon}_1+(-1)^{\tau ^a_2}\tilde{\epsilon}_6] \right)
\label{piaex2baa} 
\eea 
The required intersection numbers fix 
$\tau ^b_0=0 \bmod 2$, and then
\bea
\Pi_b^{\rm ex}&=& -{\Pi_b^{\rm ex}}' \\
&=&(-1)^{\tau ^b_0} [ {\epsilon} _1+ (-1)^{\tau ^b_2}{\epsilon}_5]  \label{bex2baa}
\eea

Similarly, for the second solution $\tau ^a_1=1 \bmod 2$ and we find
\bea
\Pi_a^{\rm ex}&=&(-1)^{\tau ^a_0 }\left( [{\epsilon}_1+(-1)^{\tau ^a_2}{\epsilon}_6] -
[\tilde{\epsilon}_1+(-1)^{\tau ^a_2}\tilde{\epsilon}_6] \right)
\label{piaex3baa} 
\eea 
which is just (minus) the orientifold dual of (\ref{piaex2baa}). 
The $SU(2)_L$ stack $b$ is the same as for the first solution and is given in (\ref{bex2baa}).

The third and fourth solutions, displayed in the bottom half of Table \ref{baa2}, have $SU(3)_{\rm colour}$ stacks $a$ 
that are the orientifold duals of the two solutions in the upper half of the table.  Thus the exceptional parts are given by 
(\ref{piaex3baa}) and (\ref{piaex2baa}) respectively. They have the same $SU(2)_L$ stack $b$, and the required intersection numbers occur when $\tau^b_1=0 \bmod 2$. Thus $\Pi _b^{\rm ex} $ is the same as that found in \S \ref{sbaa1} and given in
(\ref{bex141}).
\subsubsection{Solution with ${\rm Im} \ U_3=-2\sqrt{3}$}
The absence of symmetric representations on the $SU(3)_{\rm colour}$ stack $a$ of the 
 solution in Table \ref{baa3} requires that $\tau ^a_1=0 \bmod 2$ so that
\bea
\Pi_a^{\rm ex}&=&(-1)^{\tau ^a_0 }\left( [{\epsilon}_1+(-1)^{\tau ^a_2}{\epsilon}_5] -
[\tilde{\epsilon}_1+(-1)^{\tau ^a_2}\tilde{\epsilon}_5] \right)
\label{piaex4baa} 
\eea 
Again, the $SU(2) $ stack $b$ is the same as that found in \S \ref{sbaa1} and given in
(\ref{bex141}).
\subsubsection{Solution with ${\rm Im} \ U_3=-\sqrt{3}/2$}
The absence of symmetric representations on the $SU(3)_{\rm colour}$ stack $a$ of the 
 solution in Table \ref{baa4} requires that $\tau ^a_1=1 \bmod 2$ so that
\bea
\Pi_a^{\rm ex}&=&(-1)^{\tau ^a_0+1 }\left( [{\epsilon}_1+(-1)^{\tau ^a_2}{\epsilon}_4] -
[\tilde{\epsilon}_1+(-1)^{\tau ^a_2}\tilde{\epsilon}_4] \right)
\label{piaex5baa} 
\eea 
The $SU(2) $ stack $b$ is the same as that found in for the first two solutions in \S \ref{sbaa2} and given in
(\ref{bex2baa}).
 \subsection{ABA lattice}    \label{abal}
On the {\bf BAA} lattice the supersymmetry constraints for a general stack $\kappa$
 are the same as for the {\bf BAA} lattice given in (\ref{Xbaa}) and (\ref{Ybaa}). 
 We  found one solution having the required properties  with
\bea
{\rm Im} \ U_3 &=& -\frac{1}{2 \sqrt{3}} \label{imu3aba} 
\eea
 It is displayed in Table \ref{aba}.
\begin{table}
 \begin{center}
\begin{tabular}{||c|c|c||c|c|c||} \hline \hline
$(n^a_1,m^a_1;n^a_2,m^a_2;n^a_3,m^a_3)$&$(A^a_1,A^a_3,A^a_4,A^a_6)$&$\#({\bf A}_a)$&$(n^b_1,m^b_1;n^b_2,m^b_2;n^b_3,m^b_3)$&$(A^b_1,A^b_3,A^b_4,A^b_6)$&$\#({\bf A}_b)$  \\ \hline \hline

$(1,0;1,0;1,2)$&$(1,0,2,0)$&-3 &$(-2,1;-2,1;0,1)$&$(0,0,0,-3)$&0 \\ \hline \hline

 \end{tabular}
\end{center} 
\caption{ \label{aba} Solution on the {\bf ABA} lattice.}
 \end{table} 
From Table 5 of \cite{Bailin:2006zf} the ${\rm O6}$-plane is 
\beq
\Pi _{\rm O6}= 2\rho_1+\rho _3 -3\rho_6
\eeq
on the  {\bf ABA} lattice. 
Hence, for the solution displayed in Table \ref{aba}
\bea
 \Pi_a^{\rm bulk}\cap \Pi _{\rm O6}&=&-6  \\
\Pi_a^{\rm bulk} \cap {\Pi_a^{\rm bulk}}'&=&-8 \\
\Pi_a^{\rm ex}\cap {\Pi_a^{\rm ex}}' &=&4[1-2(-1)^{\tau ^a_1}]
\eea 
and the absence of symmetric representations  on $a$ is guaranteed provided that
\beq
\tau ^a_1=0 \bmod 2 \label{taua1aba}
\eeq
Hence, 
\bea
\Pi_a^{\rm ex}&=&(-1)^{\tau ^a_0+1 }
[\tilde{\epsilon}_1+(-1)^{\tau ^a_2}\tilde{\epsilon}_4] 
\label{piaexaba} 
\eea

As before, the absence of symmetric representations on  the $SU(2)_L$ stack $b$ is guaranteed independently of the  choice of $\tau ^b_1$. However, the required intersection numbers arise only if $\tau ^b_1=0 \bmod 2$, and then
\bea
\Pi_b^{\rm ex}&=& -{\Pi_b^{\rm ex}}' \nonumber \\
&=&(-1)^{\tau ^b_0+1} \left([ {\epsilon} _1+ (-1)^{\tau ^b_2}{\epsilon}_5] +
2 [ \tilde{\epsilon} _1+ (-1)^{\tau ^b_2}\tilde{\epsilon}_5] \right)\label{bex15}
\eea
\subsection{BBA lattice}    \label{bbal}
On the {\bf BBA} lattice the supersymmetry constraints for a general stack $\kappa$ are
\bea
X^{\kappa} &\equiv & A^{\kappa}_1+A^{\kappa}_3 +(A^{\kappa} _4-A^{\kappa}_6 )\sqrt{3}\ {\rm Im} \ U_3>0  \label{Xbba} \\
Y^{\kappa} &\equiv & \sqrt{3}(A^{\kappa}_3-A^{\kappa}_1) +(A^{\kappa} _4-A^{\kappa}_6 )\sqrt{3}\ {\rm Im} \ U_3=0
  \label{Ybba}
\eea
 We again found one solution with the required properties with
\bea
{\rm Im} \ U_3 &=& 2{ \sqrt{3}} \label{imu3bba} \\
\eea
This is displayed in Table \ref{bba}.
\begin{table}
 \begin{center}
\begin{tabular}{||c|c|c||c|c|c||} \hline \hline
$(n^a_1,m^a_1;n^a_2,m^a_2;n^a_3,m^a_3)$&$(A^a_1,A^a_3,A^a_4,A^a_6)$&$\#({\bf A}_a)$&$(n^b_1,m^b_1;n^b_2,m^b_2;n^b_3,m^b_3)$&$(A^b_1,A^b_3,A^b_4,A^b_6)$&$\#({\bf A}_b)$  \\ \hline \hline

$(0,1;0,-1;-2,1)$&$(0,2,0,-1)$&-3&$(1,-2;-2,1;1,0)$&$(3,3,0,0)$&0 \\ \hline \hline

 \end{tabular}
\end{center} 
\caption{ \label{bba} Solution on the {\bf BBA} lattice.}
 \end{table} 
From Table 5 of \cite{Bailin:2006zf} the ${\rm O6}$-plane is 
\beq
\Pi _{\rm O6}= 3 \rho_1+3 \rho _3 +\rho _4-\rho_6
\eeq
on the  {\bf BBA} lattice. 
Hence, for the solution displayed in Table \ref{bba}
\bea
 \Pi_a^{\rm bulk}\cap \Pi _{\rm O6}&=&-6  \\
\Pi_a^{\rm bulk} \cap {\Pi_a^{\rm bulk}}'&=&-8 \\
\Pi_a^{\rm ex}\cap {\Pi_a^{\rm ex}}' &=&4[1-2(-1)^{\tau ^a_1}]
\eea 
and the absence of symmetric representations  on $a$ is guaranteed provided that
\beq
\tau ^a_1=0 \bmod 2 \label{taua1bba}
\eeq
Hence, 
\bea
\Pi_a^{\rm ex}&=&(-1)^{\tau ^a_0+1 }
[{\epsilon}_1+(-1)^{\tau ^a_2}{\epsilon}_5] 
\label{piaexbba} 
\eea

The absence of symmetric representations on  the $SU(2)_L$ stack $b$ is again guaranteed independently of the  choice of $\tau ^b_1$, and the required intersection numbers arise only if $\tau ^b_1=0 \bmod 2$. Then
\bea
\Pi_b^{\rm ex}&=& -{\Pi_b^{\rm ex}}' \nonumber \\
&=&(-1)^{\tau ^b_0+1} \left([ {\epsilon} _1+ (-1)^{\tau ^b_2}{\epsilon}_4] -
 [ \tilde{\epsilon} _1+ (-1)^{\tau ^b_2}\tilde{\epsilon}_4] \right)\label{bex14bba}
\eea

\section{No-go results when $T^2_3$ is of {\bf B}-type} \label{nogot2b}
We must now  see whether it is possible to find further stacks $\kappa = c,d,...$ of fractional branes with $N_{\kappa}=1$ so that the quark- and lepton-singlet content, as well as the lepton- and Higgs-doublet content of the standard model arises at intersections of $a$ and $b$ with these new stacks,  and/or at intersections of the new stacks with each other, and/or, for the singlet-matter, as symmetric representations on the some or all of the stacks. 

At the $a \cap \kappa$ intersections of the $SU(3)_{\rm colour}$ stack $a$ with a $U(1)$ stack $\kappa$ there is chiral matter in the $({\bf 3}, {\bf 1})$ representation of $SU(3)_{\rm colour} \times SU(2)_L$, which { must} correspond to quark-singlet matter if we are to get just the standard-model spectrum. It follows from (\ref{Y}) and (\ref{ya}) that the weak hypercharge of such matter is $Y= \frac{1}{6}-y_{\kappa}$. Thus if $y_{\kappa}=\frac{1}{2}$, the colour-triplet matter will be $d$-quark singlets, while if $y_{\kappa}=-\frac{1}{2}$, it will be $u$-quark singlets; no other values of $y_{\kappa}$ are permitted if we insist on the standard-model spectrum. Likewise, at the $a \cap \kappa'$ intersections of $a$ with the orientifold image $\kappa'$ of $\kappa$, there will be colour-triplet $u$-quark singlet matter if $y_{\kappa}=\frac{1}{2}$, and $d$-quark singlet matter if $y_{\kappa}=-\frac{1}{2}$. For our purposes, we require that
\beq
-3 \leq a \cap \kappa,  a \cap \kappa' \leq 0 \label{akapineq}
\eeq
corresponding to not more than three $d^c_L$ or $u^c_L$ states. 

Similarly, using (\ref{yb}), at the intersections of the $SU(2)_L$ stack $b$ with  $\kappa$ and $\kappa'$ there is chiral matter in the $({\bf 1}, {\bf 2})$ representation of $SU(3)_{\rm colour} \times SU(2)_L$ with $Y=-y_{\kappa}$ and $Y=y_{\kappa}$ respectively. If $y_{\kappa}=\frac{1}{2}$,  the former corresponds to lepton $L$ or Higgs $\bar{H}$ doublets and the latter to $H$ doublets, and vice versa if  $y_{\kappa}=-\frac{1}{2}$. For the standard-model spectrum we require that there are four doublets with $Y=-\frac{1}{2}$ and one with $Y=\frac{1}{2}$, and this requires that for at least one stack $\kappa$
\beq
b \cap \kappa - b \cap \kappa' = 1 \bmod 2 \label{bkbk1}
\eeq
In many cases it turns out that this is a very restrictive constraint.

In all of the solutions presented in the last section (for lattices in which $T^2_3$ is of {\bf A}-type) the $SU(2)_L$ stack $b$  has the property that 
\bea
\Pi_b^{\rm bulk}&=&{\Pi_b^{\rm bulk}}' \label{bb1bulk} \\
\Pi_b^{\rm ex}&=&-{\Pi_b^{\rm ex}}' \label{bb1ex}
\eea
and the same is true for some of the solutions presented in \cite{Bailin:2006zf} for lattices in which $T^2_3$ is of {\bf B}-type. For the solutions of which this is true it follows that, for any stack $\kappa$, and in particular any of the $U(1)$ stacks $\kappa=c,d,...$
\bea
b \cap \kappa- b \cap \kappa '&=& \frac{1}{2}\Pi_b^{\rm bulk} \cap \Pi_{\kappa}^{\rm bulk} \label{bkapdiff} \\
b \cap \kappa+ b \cap \kappa '&=& \frac{1}{2}\Pi_b^{\rm ex} \cap \Pi_{\kappa}^{\rm ex} \label{bkapsum}
\eea
\subsection{{\bf BBB} and {\bf ABB} lattices} \label{bbb}
To see how restrictive (\ref{bkbk1}) is, we consider  the   solutions presented in \S 7.4 of \cite{Bailin:2006zf} for the {\bf BBB} lattice. All three solutions have the same $SU(3)_{\rm colour}$ stack $a$ (denoted by $b$ in \cite{Bailin:2006zf}):
\bea
\Pi_a^{\rm bulk}&=&\rho _1-\rho_4 \label{abbb} \\
{\Pi_a^{\rm ex}}&=&(-1)^{\tau ^a_0}[\tilde {\epsilon} _1+ (-1)^{\tau ^a_2}\tilde{\epsilon}_6]  \label{aexbbb16} \\
{\rm or} \ &=& (-1)^{\tau ^a_0}[\tilde {\epsilon} _4+ (-1)^{\tau ^a_2}\tilde{\epsilon}_5]  \label{aexbbb15}
\eea
In the first solution the $SU(2)_L$ stack $b$ has 
\bea
\Pi_b^{\rm bulk}&=&\rho_1-\rho_3 -2\rho_4+2\rho_6={\Pi_b^{\rm bulk}}' \label{bbbb} \\
\Pi_b^{\rm ex}&=&-{\Pi_b^{\rm ex}}' \\
&=&(-1)^{\tau ^b_0}\left([ \epsilon _1+ (-1)^{\tau ^b_2}\epsilon_4]-[ \tilde{\epsilon} _1+ (-1)^{\tau ^b_2}\tilde{\epsilon}_4]  \right) \label{bexbbb14}\\
{\rm or} \ &=& (-1)^{\tau ^b_0} \left([  \epsilon _5+ \epsilon_6]-[\tilde{\epsilon} _5+ \tilde{\epsilon}_6]\right)  \label{bexbbb56}
\eea
The supersymmetry constraint on this lattice
\beq
Y^b \equiv \sqrt{3} ( A^b_3-A^b_1+\frac{1}{2}A^b_6-\frac{1}{2}A^b_4)+(A^b_4+A^b_6) {\rm Im} \ U_3=0 \label{Ybbbb}
\eeq
then requires that
\beq
{\rm Im} \ U_3=-\frac{\sqrt{3}}{2}
\eeq
so that all stacks $\kappa$ are required to satisfy
\bea
X^{\kappa} &\equiv& A^{\kappa}_1+A^{\kappa}_3-A^{\kappa}_4+2A^{\kappa}_6>0 \label{Xkbbb}\\
\frac{1}{\sqrt{3}}Y^{\kappa} &\equiv& A^{\kappa}_3-A^{\kappa}_1-A^{\kappa}_4=0 \label{Ykbbb}
\eea
It is easy to see that this requires that the bulk wrapping numbers $A^{\kappa}_p$ satisfy
\bea
(A^{\kappa}_1,A^{\kappa}_3,A^{\kappa}_4,A^{\kappa}_6)&=& (1,1,0,0) \bmod 2 \label{cA}\\
{\rm or} \ &=& (1,0,1,0) \bmod 2 \label{dA}\\
{\rm or} \ &=& (0,0,0,1) \bmod 2 
\eea
and hence that the wrapping numbers $(n^{\kappa}_k, m^{\kappa}_k)$ on the torus $T^2_k$ satisfy
\bea
(n^{\kappa}_1, m^{\kappa}_1;n^{\kappa}_2, m^{\kappa}_2;n^{\kappa}_3, m^{\kappa}_3) &=&(1,0;0,1;1,0) \bmod 2 \label{cnm}\\
{\rm or} \ &=&(1,1;0,1;1,1) \bmod 2  \label{dnm} \\
{\rm or} \ &=&(0,1;0,1;0,1) \bmod 2 
\eea
respectively, when we choose the representative 3-cycle in which $(n^{\kappa}_1, m^{\kappa}_1)=(n^{\kappa}_3, m^{\kappa}_3) \bmod 2$. In fact, the last two cases are interchanged under the action of $\mathcal{R}$, so that we need only consider the first two possibilities. We denote by $c$ stacks with wrapping numbers satisfying (\ref{cA}), (\ref{cnm}), and by $d$ stacks with wrapping numbers satisfying (\ref{dA}), (\ref{dnm}). 
It follows from (\ref{bkapdiff}) and (\ref{bbbb}) that for this solution
\bea
b \cap \kappa - b \cap \kappa '&=&3(2A^{\kappa}_1-2A^{\kappa}_3+A^{\kappa}_4-A^{\kappa}_6) \\
&=&-3(A^{\kappa}_4+A^{\kappa}_6) \label{bkdiff}
\eea
using (\ref{Ykbbb}), which only satisfies (\ref{bkbk1}) if $\kappa$ is of type $d$. Further, the only solutions that do not entail unwanted vector-like doublets have
\beq
A^{d}_4+A^{d}_6= \epsilon
\eeq
where $\epsilon = \pm 1$. It is easy to see that the only solution consistent with supersymmetry (\ref{Xkbbb}),(\ref{Ykbbb}) and the requirement that
\beq
A^{\kappa}_1A^{\kappa}_6=A^{\kappa}_3A^{\kappa}_4
\eeq 
is when $\epsilon=-1$ and
\beq
(A^{d}_1,A^{d}_3,A^{d}_4,A^{d}_6)=(1,0,-1,0)
\eeq
Then the wrapping numbers  are given by
\beq
(n^{d}_1, m^{d}_1;n^{d}_2, m^{d}_2;n^{d}_3, m^{d}_3)=(\eta \chi,-\eta \chi;0,\chi;\eta, -\eta) \label{ndkmdk}
\eeq
where $\eta, \chi = \pm 1$. Using the results presented in \cite{Bailin:2006zf}, the general form for the exceptional part of  a $d$-type stack is given by 
\bea
\Pi _d^{\rm ex}&=&(-1)^{\tau ^d_0} \left( [m^d_2 - (-1)^{\tau ^d_1}(n^d_2+m^d_2)][ \epsilon _1+ (-1)^{\tau ^d_2}\epsilon_6] +
[n^d_2 +m^d_2- (-1)^{\tau ^d_1}n^d_2][\tilde {\epsilon} _1+ (-1)^{\tau ^d_2}\tilde{\epsilon}_6]  \right) \nonumber \\
&& \label{dex16} \\
{\rm or} \ &=& (-1)^{\tau ^d_0} \left( [m^d_2 - (-1)^{\tau ^d_1}(n^d_2+m^d_2)][ \epsilon _4+ (-1)^{\tau ^d_2}\epsilon_5] +
[n^d_2 +m^d_2- (-1)^{\tau ^d_1}n^d_2][\tilde {\epsilon} _4+ (-1)^{\tau ^d_2}\tilde{\epsilon}_5]  \right) \nonumber\\
&& \label{dex45}
\eea
 Then, using (\ref{ndkmdk}), it follows  from  (\ref{bkapsum}) and (\ref{bexbbb14}) or (\ref{bexbbb56}) that 
\beq
b \cap d + b \cap d'=\pm 1 \ {\rm or} \ \pm 3
\eeq
and hence that
\beq
(b \cap d, b \cap d')=(1,-2),\quad (2,-1), \quad (3,0) \quad {\rm or} \quad (0,-3)
\eeq
In all four cases such a stack will give three  $L$ or $\bar{H}$ doublets with $Y=-\frac{1}{2}$ provided that we choose $y_d=-\frac{1}{2}$. 
However, there remains to be found a pair of doublets with $Y=\frac{1}{2},-\frac{1}{2}$ arising at intersections of the stack $b$ with a different  $U(1)$ stack $\kappa$ satisfying
\beq
(b \cap \kappa, b \cap \kappa')=\pm(1,1) \label{bk11}
\eeq
(There is no possibility of utilising {\em two} further stacks, since it follows from (\ref{bkdiff}) that the only solutions satisfying (\ref{bkbk1}) necessarily have $b \cap \kappa - b \cap \kappa '  = 0 \bmod 3$.)
It follows from (\ref{bkdiff}) that $\kappa$ must be of type $c$. Now, the general form for the exceptional part of  a $c$-type stack is given by 
\bea
\Pi _c^{\rm ex}&=&(-1)^{\tau ^c_0} \left( [-(n^c_2+m^c_2) + (-1)^{\tau ^c_1}n^c_2][ \epsilon _1+ (-1)^{\tau ^c_2}\epsilon_4] -[n^c_2 + (-1)^{\tau ^c_1}m^c_2][\tilde {\epsilon} _1+ (-1)^{\tau ^c_2}\tilde{\epsilon}_4]  \right)  \nonumber \\
&& \label{cex14} \\
 {\rm or} \ &=&(-1)^{\tau ^c_0} \left( [-(n^c_2+m^c_2) + (-1)^{\tau ^c_1}n^c_2][ \epsilon _5+ (-1)^{\tau ^c_2}\epsilon_6] -[n^c_2 + (-1)^{\tau ^c_1}m^c_2][\tilde {\epsilon} _5+ (-1)^{\tau ^c_2}\tilde{\epsilon}_6]  \right)  \nonumber \\
&& \label{cex56}
\eea
Using (\ref{bkapsum}) and (\ref{bexbbb14}) or (\ref{bexbbb56}) this gives
\bea
b \cap c + b \cap c'&=&(-1)^{\tau ^b_0+\tau ^c_0+1}[1+(-1)^{\tau ^a_2+\tau ^c_2}]\left(m^c_2-(-1)^{\tau ^c_1}(n^c_2+m^c_2) \right) \\
&=&0 \bmod 4
\eea
since $(n^c_2,m^c_2)=(0,1) \bmod 2$. It follows that we cannot satisfy (\ref{bk11}) using a type $c$ stack, and certainly not using $d$-type. We conclude that this solution can{\em not} produce just the standard-model spectrum. A similar argument shows that the other solutions on the {\bf BBB} lattice also cannot yield the required doublet spectrum. In fact, the same conclusion, reached by a similar argument, also holds for the three solutions found on the {\bf ABB} lattice. 
\subsection{{\bf AAB}  and {\bf BAB} lattices} \label{aabbab}
For the other two lattices in which $T^2_3$ is of {\bf B}-type the situation is different. On each of these lattices there is one solution with the property (\ref{bb1bulk}) and (\ref{bb1ex}), and an argument similar to that given above shows that they too cannot yield the required doublet spectrum. For the other solution on each of these two lattices
 it is the $SU(3)_{\rm colour}$ stack $a$ that has the property
\bea
\Pi_a^{\rm bulk}&=&{\Pi_a^{\rm bulk}}' \label{aa1bulk}\\
\Pi_a^{\rm ex}&=&-{\Pi_a^{\rm ex}}' \label{aa1ex}
\eea
so that 
\bea
a \cap \kappa- a \cap \kappa '&=& \frac{1}{2}\Pi_a^{\rm bulk} \cap \Pi_{\kappa}^{\rm bulk} \label{akapdiff} \\
a \cap \kappa+ a \cap \kappa '&=& \frac{1}{2}\Pi_a^{\rm ex} \cap \Pi_{\kappa}^{\rm ex} \label{akapsum}
\eea
This is the case for the first solution on the {\bf AAB} lattice, given in \S 7.1 of \cite{Bailin:2006zf}, in which  
\bea
\Pi_a^{\rm bulk}&=&\rho _1 ={\Pi_a^{\rm bulk}}'\label{aaab} \\
{\Pi_a^{\rm ex}}&=&-{\Pi_a^{\rm ex}}' \\
&=& (-1)^{\tau ^a_0+1} \left(2[ \epsilon _5 -\epsilon _6]+
[\tilde {\epsilon} _5-\tilde{\epsilon}_6] \right) \label{aexaab56}
\eea
and 
\bea
\Pi_b^{\rm bulk}&=&\rho_1+\rho_3 -\rho_4-\rho_6  \label{baab} \\
\Pi_b^{\rm ex}&=&(-1)^{\tau ^b_0+1}\left([ \epsilon _1+ (-1)^{\tau ^b_2}\epsilon_6]-[ \tilde{\epsilon} _1(-1)^{\tau ^b_2}\tilde{\epsilon}_6]  \right) \label{bexaab16}\\
{\rm or} \ &=& (-1)^{\tau ^b_0} \left([  \epsilon _4+ (-1)^{\tau ^b_2}\epsilon_5]-[\tilde{\epsilon} _4- (-1)^{\tau ^b_2}\tilde{\epsilon}_5]  \right) \label{bexaab45}
\eea
The supersymmetry constraint on this lattice
\beq
Y^b \equiv \sqrt{3} ( A^b_3+\frac{1}{2}A^b_6)+(2A^b_4-A^b_6) {\rm Im} \ U_3=0 \label{Ybaab}
\eeq
then requires that
\beq
{\rm Im} \ U_3=\frac{\sqrt{3}}{2}
\eeq
so that all stacks $\kappa$ are required to satisfy
\bea
X^{\kappa} &\equiv& 2A^{\kappa}_1-A^{\kappa}_3+A^{\kappa}_4-2A^{\kappa}_6>0 \label{Xkaab}\\
\frac{1}{\sqrt{3}}Y^{\kappa} &\equiv& A^{\kappa}_3+A^{\kappa}_4=0 \label{Ykaab}
\eea
This requires that the bulk wrapping numbers $A^{\kappa}_p$ satisfy
\bea
(A^{\kappa}_1,A^{\kappa}_3,A^{\kappa}_4,A^{\kappa}_6)&=& (1,0,0,0) \bmod 2 \label{cA2}\\
{\rm or} \ &=& (1,1,1,1) \label{dA2} \bmod 2 \\
{\rm or} \ &=& (0,0,0,1) \bmod 2 
\eea
and hence that the wrapping numbers $(n^{\kappa}_k, m^{\kappa}_k)$ on the torus $T^2_k$ satisfy
\bea
(n^{\kappa}_1, m^{\kappa}_1;n^{\kappa}_2, m^{\kappa}_2;n^{\kappa}_3, m^{\kappa}_3) &=&(1,0;1,0;1,0) \bmod 2 \label{cnm2}\\
{\rm or} \ &=&(1,1;1,1;1,1) \bmod 2  \label{dnm2} \\
{\rm or} \ &=&(0,1;0,1,0,1) \bmod 2 
\eea
respectively, when we choose the representative 3-cycle in which $(n^{\kappa}_1, m^{\kappa}_1)=(n^{\kappa}_3, m^{\kappa}_3) \bmod 2$. The last two cases are interchanged under the action of $\mathcal{R}$, so that we need only consider the first and second possibilities. We denote by $c$ stacks with wrapping numbers satisfying (\ref{cA2}), (\ref{cnm2}), and by $d$ stacks with wrapping numbers satisfying (\ref{dA2}), (\ref{dnm2}).
As noted earlier, the intersections of the $SU(3)_{\rm colour}$ stack $a$ with a $U(1)$ stack $\kappa$ produce quark-singlet matter. It is obvious from (\ref{aaab}) and (\ref{aexaab56}),  using (\ref{noAa}),  that 
\beq
a \cap a' = 0=\#({\bf A}_a)
\eeq
Since there is no antisymmetric matter on $a$,
  all of the quark-singlet matter must arise  at  intersections of $a$ with such $U(1)$ stacks. Using (\ref{aaab}), (\ref{akapdiff}), (\ref{aexaab56}) and (\ref{akapsum}) we find that
\beq
a \cap c- a \cap c'=2A^{c}_4-A^{c}_6=0 \bmod 4 
\eeq
The last equality follows because $A^c_1A^c_6=A^c_3A^c_4=0 \bmod 4$ for a $c$-type stack.
The general form for the exceptional part of  a $c$-type stack is given by (\ref{cex14}) or (\ref{cex56}). Then, 
using (\ref{akapsum}) and (\ref{aexaab56}) or (\ref{aexaab56}), it follows that
\bea
a \cap c+a \cap c'&=&(-1)^{\tau ^a_0+\tau ^c_0}[1+(-1)^{\tau ^a_2+\tau ^c_2}][m^c_2-n^c_2-(-1)^{\tau ^c_1}(n^c_2+2m^c_2)] \\
&=& 0 \bmod 4
\eea
Thus, the only solutions satisfying (\ref{akapineq}) are
\beq
(a \cap c, a \cap c')=(0,0) \  {\rm and} \ (-2,-2)
\eeq
Similarly, we find that
\bea
a \cap d- a \cap d'&=&2A^{d}_4-A^{d}_6=1 \bmod 2 \\
a \cap d+a \cap d'&=&(-1)^{\tau ^a_0+\tau ^d_0}[2n^d_2+m^c_2-(-1)^{\tau ^d_1}(n^d_2-m^d_2)]= 1 \bmod 2
\eea
which, in principle, allows
\beq
(a \cap d, a \cap d')=(\underline{-1,0}), \ (\underline{-1,-2}), \ (\underline{-3,-2}), \ {\rm and} \ (\underline{-3,0}) \label{adposs}
\eeq
where the underlining signifies that either ordering is allowed.
Clearly we cannot obtain the required quark-singlet content without using at least two $d$-type stacks. The first three of the above four possibilities require that 
\beq
2A^{d}_4-A^{d}_6= \epsilon
\eeq
where $\epsilon= \pm 1$. The only solution consistent with supersymmetry (\ref{Xkaab}),(\ref{Ykaab}) and the requirement that
\beq
A^{d}_1A^{d}_6=A^{d}_3A^{d}_4
\eeq 
is when $\epsilon=-1$ and
 \beq
(A^{d}_1,A^{d}_3,A^{d}_4,A^{d}_6)=(1,1,-1,-1)
\eeq
But then
\beq
(a \cap d, a \cap d') \neq ({-1,0}), \ ({-2,-1}),  \ {\rm or} \ ({-3,-2})
\eeq
A similar argument for the fourth possibility in (\ref{adposs}) shows that
\beq
(a \cap d, a \cap d') \neq  ({-3,0})
\eeq
In all cases, therefore, $a \cap d > a \cap d'$, and we are unable to  achieve equal numbers of $u^c_L$ and $d^c_L$ quark singlet states using only $d$-type stacks. We noted above that  only $c$-type stacks with $a \cap c=a \cap c'= 0 \bmod 2$ are allowed, so it follows that we can{\em not} obtain just the standard-model quark-singlet spectrum from this model. A similar argument  applies to the solution on the {\bf BAB} lattice which also has the property (\ref{aa1bulk}) and (\ref{aa1ex}). We conclude that for one reason or the other none of the solutions presented in \cite{Bailin:2006zf}, in which $T^2_3$ is of {\bf B}-type, can yield just the standard model spectrum.


\section{Solutions for {\bf AAA} lattice} \label{aaa}
Fortunately, some, but not all,  of the solutions on lattices in which $T^2_£$ is of {\bf A}-type {\em can} be extended to give the standard-model spectrum.
\subsection{Solutions with ${\rm Im} \ U_3=-1/\sqrt{3}$} \label{ssaaa1}
Consider first the  solutions on the {\bf AAA} lattice with ${\rm Im} \ U_3=-1/\sqrt{3}$ presented in  Table \ref{aaa1}. 
On this lattice the supersymmetry constraint (\ref{Yaaa}) requires that the bulk wrapping numbers for all stacks $\kappa$ 
satisfy 
\bea
(A^{\kappa}_1,A^{\kappa}_3,A^{\kappa}_4,A^{\kappa}_6) &=& (1,0,0,0) \bmod 2 \label{cAAA}\\
{\rm or} \ &=& (1,\theta,1,\theta) \bmod 2 \label{dAAA} \\
{\rm or} \ &=& (0,0,1,0) \bmod 2 \label{eAAA} 
\eea
where $\theta=0,1$. 
This restricts the allowed wrapping numbers $(n^{\kappa}_k,m^{\kappa}_k) $ to the cases
 \bea
 (n^{\kappa}_1,m^{\kappa}_1;n^{\kappa}_2,m^{\kappa}_2;n^{\kappa}_3,m^{\kappa}_3)&=& (1,0;1,0;1,0) \bmod 2  \label{cnmAAA}\\
 {\rm or} \ &=& (1,1;\theta,1;1,1) \bmod 2 \label{dnmAAA}  \\
{\rm or} \ &=& (0,1;1,1;0,1) \bmod 2 \label{enmAAA} 
\eea
respectively, in the ``gauge'' in which $(n^{\kappa}_1,m^{\kappa}_1)=(n^{\kappa}_3,m^{\kappa}_3) \bmod 2$. 
As before, there is also a fourth class which may be obtained by the action of $\cal{R}$ on (\ref{dAAA}) and (\ref{dnmAAA}), but we need only consider one of them.
We denote by $c$ stacks with wrapping numbers satisfying (\ref{cAAA}), (\ref{cnmAAA}),  by $d_{\theta}$ stacks with wrapping numbers satisfying (\ref{dAAA}), (\ref{dnmAAA}), and by $e$ stacks with wrapping numbers satisfying (\ref{eAAA}), (\ref{enmAAA}). 
The three solutions in the lower half of the Table all have the same $SU(2)_L$ stack $b$ with
\beq
\Pi _b^{\rm bulk}=\rho _4+2\rho _6={\Pi _b^{\rm bulk}}'
\eeq
and $\Pi _b^{\rm ex}$ given in (\ref{pibx1}) and (\ref{pibex456}). 
It follows from (\ref{bkapdiff})  that for this solution
\bea
b \cap \kappa - b \cap \kappa '&=&-3A^{\kappa}_3 \label{bkapaaa} \\
&=&A^{\kappa}_6-2A^{\kappa}_4 \label{bkapaaa2}
\eea
the last line following from the supersymmetry constraint (\ref{Yaaa}). This only 
  satisfies (\ref{bkbk1}) if $\kappa$ is of type $d_1$.
Further,  the only possibility that does not entail unwanted vector-like doublets is when
\beq
A^{d_1}_3= \epsilon  \label{ad46}
\eeq
where $\epsilon = \pm 1$. The only solutions satisfying the supersymmery constraints (\ref{Xaaa}), (\ref{Yaaa})
and the consistency condition, namely
\beq
A^{d_1}_1A^{d_1}_6=A^{d_1}_3A^{d_1}_4 \label{Ad1634}
\eeq
are 
\bea
(A^d_1,A^d_3,A^d_4,A^d_6)&=&(1,1,3,3) \label{Adp1133} \\
{\rm or} \ &=&(1,-1,-1,1) \label{Adp1111}
\eea
The former  requires that
\beq
(n^d_1,m^d_1)= \eta \chi(-1,1), \ (n^d_2,m^d_2)=  \chi(1,-1), \ (n^d_3,m^d_3)= \eta (1,
3) \label{nkdmkd}
\eeq 
Now, from (\ref{pibex}), we have that
\beq
b \cap \kappa + b \cap \kappa '= \frac{1}{2} \Pi _b^{\rm ex} \cap \Pi _{\kappa}^{\rm ex}
\eeq
Then (\ref{nkdmkd}) gives
\beq
(b \cap d , b \cap d')=(-2,1) \ {\rm or} \ (-1,2) \label{bdbd1n}
\eeq
Hence, we must take 
\beq
y_d=-\frac{1}{2} \label{yd1133}
\eeq
so that there are three lepton  $L$ or Higgs $\bar{H}$ doublets with weak hypercharge $Y=-\frac{1}{2}$.  The latter possibility (\ref{Adp1111}) requires that
\bea
&&(n^d_1,m^d_1)= \eta \chi(1,-1), \ (n^d_2,m^d_2)=  \chi(1,1), \ (n^d_3,m^d_3)= \eta (1,-1) \label{nkdmkd11111}
\\
&{\rm or}& \ (n^d_1,m^d_1)= \eta \chi(1,1), \ (n^d_2,m^d_2)=  \chi(1,-1), \ (n^d_3,m^d_3)= \eta (1,-1) \label{nkdmkd11112}
\eea
These give
\bea
(b \cap d , b \cap d')&=&(0,-3) \ {\rm or} \ (3,0) \\
{\rm or} \ &=&(2,-1) \ {\rm or} \ (1,-2)
\eea
respectively, and in this case  we must take
\beq
y_d=\frac{1}{2} \label{yd1111}
\eeq

Both possibilities require that there is precisely {\em one} further $U(1)$ stack $\kappa$ whose intersections with $b$ give the remaining lepton/Higgs doublets
\beq
(b \cap \kappa, b \cap \kappa') = \pm (1,1) \label{bkap11}
\eeq
Then, from (\ref{bkapaaa}), (\ref{bkapaaa2}) and the consistency condition, we infer that $\kappa$ must be of type $e$ or of type $c$ with
\bea
(A^e_1,A^e_3,A^e_4,A^e_6)&=&(2j+1)(0,0,1,2) \label{Ae0012} \\
{\rm or} \ (A^c_1,A^c_3,A^c_4,A^c_6) &=&(2j+1)(1,0,0,0) \label{Ac1000}
\eea
with the integer $j \geq 0$ to ensure positivity of $X^{e,c}$. 
However,
\bea
\Pi _b ^{\rm ex} \cap \Pi _e^{\rm ex}&=&(-1)^{\tau ^b_0+\tau ^e _0 +1}2[1+(-1)^{\tau ^b _2+\tau ^e_2}][m^e_2 + (-1)^{\tau ^e_1}n^e_2] \\
&=& 0 \bmod 8
\eea
since an $e$ stack has $(n^e_2,m^e_2)=(1,1) \bmod2$. Hence
\bea
b \cap e + b \cap e'&=&\frac{1}{2}\Pi _b ^{\rm ex} \cap \Pi _e^{\rm ex}\\
&=&  0 \bmod 4
\eea
It follows that we can never satisfy  (\ref{bkap11}) with a $e$ type stack.

For a $c$-type stack, 
it follows from (\ref{Ac1000}) that 
\beq
(n^c_3,m^c_3)=\eta(1,0)
\eeq
where $\eta = \pm 1$, and that 
\beq
\left( \begin{array}{c}
             n^c_1 \\
             m^c_1
       \end{array} \right) = \frac{ \eta(2j+1)}{{n^c_2}^2+n^c_2m^c_2+{m^c_2}^2} \left( \begin{array}{c}
                                               n^c_2+m^c_2 \\
                                               -m^c_2
                                               \end{array} \right) \label{nc21mc21}
\eeq
We also require that $b \cap c + b \cap c'=\pm 2$. Hence
\beq
-(n^c_2+m^c_2)+(-1)^{\tau ^c_1}n^c_2 = 2 \phi
\eeq
where $\phi = \pm 1$. We define 
\beq
p_c \equiv n^c_2-(-1)^{\tau ^c_1}\phi= 0 \bmod 2
\eeq
Then for every pair of (coprime) wrapping numbers of the form
\beq
(n^c_2,m^c_2)=[ (-1)^{\tau ^c_1}\phi + p_c, p_c((-1)^{\tau ^c_1}-1)-\phi((-1)^{\tau ^c_1}+1)] \label{nc2mc22} 
\eeq
we are guaranteed to generate the required bulk wrapping numbers (\ref{Ac1000}) and intersection numbers
 provided that we choose $j$ such that
\beq
\left( \begin{array}{c}
             n^c_1 \\
             m^c_1
       \end{array} \right) = \frac{ \eta(2j+1)}{ [2-(-1)^{\tau ^c_1}]p_c^2+[2+(-1)^{\tau ^c_1}]}\left( \begin{array}{c}
                                              (-1)^{\tau ^c_1}p_c -\phi \\
                                               p_c[1-(-1)^{\tau ^c_1}]+[1+(-1)^{\tau ^c_1}]\phi
                                               \end{array} \right) \label{n1m1c2}
\eeq
are  also coprime integers. 
For example, if $p_c=0$, then
\beq
\left( \begin{array}{c}
             n^c_1 \\
             m^c_1
       \end{array} \right) =  \eta \phi \frac{(2j+1)}{2+(-1)^{\tau ^c_1}}\left( \begin{array}{c}
                                             -1\\
                                             (-1)^{\tau ^c_1}+1
                                               \end{array} \right)
                                               \eeq
Thus,  we must choose $j=\frac{1}{2}[1+(-1)^{\tau ^c_1}]$. In the cases that $\tau ^c_1=0 \bmod 2$ 
\beq
\left( \begin{array}{c}
             n^c_1 \\
             m^c_1
       \end{array} \right) =  \eta \phi \frac{(2j+1)}{p_c^2+3}\left( \begin{array}{c}
                                             p_c-\phi\\
                                             2 \phi
                                               \end{array} \right)
                                               \eeq
and we must choose $j=1+\frac{1}{2}p_c^2$.
We conclude that to get the required standard-model doublet spectrum we must have both a $d$-type stack, and a $c$-type stack.

Since there are no antisymmetric representations ${\bf A}_a=\bar{\bf 3}$ on the $SU(3)_{\rm colour}$ stack $a$, all of the quark singlets must arise at intersections of $a$ with the $U(1)$ stacks $d$ and $c$, unless we introduce further stacks  that have no intersections with the $SU(2)_L$ stack $b$. Consider the first of the three solutions in the bottom half of Table \ref{aaa1}. For the first possibility for $d$,  given in (\ref{Adp1133}), we have
\bea 
\Pi _a ^{\rm bulk} \cap \Pi _d ^{\rm bulk}&=&0 \\
\Pi _a ^{\rm bulk} \cap {\Pi _d ^{\rm bulk}}'&=&12
\eea
and 
\bea 
\Pi _a ^{\rm ex} \cap \Pi _d ^{\rm ex}&=&(-1)^{\tau^a _0+\tau ^d_0}2\chi[1+(-1)^{\tau ^a_2+\tau ^d_2}][1+(-1)^{\tau ^d_1}] 
\label{adx1133}\\
\Pi _a ^{\rm ex} \cap {\Pi _d ^{\rm ex}}'&=&(-1)^{\tau^a _0+\tau ^d_0}2\chi[1+(-1)^{\tau ^a_2+\tau ^d_2}][2-(-1)^{\tau ^d_1}] 
\label{ad1x1133}
\eea
Hence if 
\beq
\tau ^a_2+\tau ^d_2= 1 \bmod 2
\eeq
then 
\beq
(a \cap d, a \cap d')=(0,3)
\eeq
and, using (\ref{yd1133}), we get three $\bar{d}^c_L$ quark-singlet states from intersections of $a$ with $d'$. 
To avoid vectorlike quark-single matter, therefore, we must take
\beq
\tau ^a_2+\tau ^d_2= 0 \bmod 2
\eeq
Then if
\beq
\tau ^d_1=0 \bmod 2
\eeq
then 
\beq
(a \cap d, a \cap d')=(-2,2) \ {\rm or} \ (2,4)
\eeq
both of which violate the inequalities (\ref{akapineq}) and are therefore unacceptable.  Alternatively, if 
\beq
\tau ^d_1=1 \bmod 2
\eeq
then 
\beq
(a \cap d, a \cap d')=(0,6) \ {\rm or} \ (0,0)
\eeq
Only the latter does not violate (\ref{akapineq}), and this occurs when
\beq
(-1)^{\tau^a _0+\tau ^d_0}\chi=-1
\eeq
in which case 
\beq
\Pi _d^{\rm ex}=\Pi _a^{\rm ex}
\eeq
with the latter given in (\ref{aex4aaa}). Then $d=a$, which is unacceptable.

In the second possibility for $d$,  given in (\ref{Adp1111}), 
\bea 
\Pi _a ^{\rm bulk} \cap \Pi _d ^{\rm bulk}&=&0 \\
\Pi _a ^{\rm bulk} \cap {\Pi _d ^{\rm bulk}}'&=&-12
\eea
Using (\ref{nkdmkd11111}) we get
\bea 
\Pi _a ^{\rm ex} \cap \Pi _d ^{\rm ex}&=&(-1)^{\tau^a _0+\tau ^d_0}2\chi[1+(-1)^{\tau ^a_2+\tau ^d_2}]3[(-1)^{\tau ^d_1}-1] \\
\Pi _a ^{\rm ex} \cap {\Pi _d ^{\rm ex}}'&=&(-1)^{\tau^a _0+\tau ^d_0}2\chi[1+(-1)^{\tau ^a_2+\tau ^d_2}]3(-1)^{\tau ^d_1}
\eea
whereas using (\ref{nkdmkd11112}) we get (\ref{adx1133}) and (\ref{ad1x1133}) again.
Hence if 
\beq
\tau ^a_2+\tau ^d_2= 1 \bmod 2
\eeq
then 
\beq
(a \cap d, a \cap d')=(0,-3) \label{adad103}
\eeq
and, using (\ref{yd1111}), we get   $3{u}^c_L$ quark-singlet states from intersections of $a$ with $d'$ .
Alternatively, if 
\beq
\tau ^a_2+\tau ^d_2= 0 \bmod 2 \label{tad2}
\eeq
the only  solution  that does not entail vector-like quark-singlet matter is
\beq
(a \cap d, a \cap d')=(0,0)
\eeq
which occurs when $\tau ^d_1=0 \bmod 2$ in the case of (\ref{nkdmkd11111}) and when $\tau ^d_1=1 \bmod 2$ for (\ref{nkdmkd11112}). In both cases
\beq
(-1)^{\tau ^a_0+\tau ^d_0} \chi=+1 \label{tad0chi}
\eeq
and
\beq
\Pi _d^{\rm ex}=-\Pi _a^{\rm ex} \label{adex}
\eeq

Similarly, with the fourth stack  of type $c$, as given in (\ref{Ac1000}) and (\ref{nc2mc22}), then
\bea 
a \cap c &=&-\frac{3}{2}(2j+1)+(-1)^{\tau^a _0+\tau ^c_0+1}\frac{1}{2}[p_c((-1)^{\tau ^c_1}-2)+3 \phi]  \label{acI.1}\\
a \cap c' &=&-\frac{3}{2}(2j+1)+(-1)^{\tau^a _0+\tau ^c_0+1}\frac{1}{2}[p_c(2-(-1)^{\tau ^c_1})+3 \phi] \label{ac1I.1}
\eea
Hence, 
\bea
a \cap c -a \cap c'&=&p_c(-1)^{\tau^a _0+\tau ^c_0+1}[(-1)^{\tau ^c_1}-2] \\
&=& 0 \bmod 2
\eea
If we use the possibility (\ref{adad103}), it is therefore impossible to obtain the required $3d^c_L$ quark singlets that are needed  just from the intersections of the $SU(3)_{\rm colour}$ stack $a$ with  $c$ and $c'$.
To avoid vectorlike matter we must take
$p_c=0$ or else $|p_c|=2$ but then only  with $\tau ^c_1=0 \bmod 2$. For the former,
\beq
2j+1=2+(-1)^{\tau ^c_1}
\eeq
so that 
\bea
a \cap c= a \cap c'&=& -\frac{3}{2}[2+(-1)^{\tau ^c_1}+(-1)^{\tau^a _0+\tau ^c_0}\phi] \\
&=&0, \  -3 \ {\rm or} \ -6
\eea
For either choice of $y_c= \pm \frac{1}{2}$, we get  all six  quark-singlet states $3u^c_L+3 d^c_L$, when
\beq
(-1)^{\tau ^c_1}+(-1)^{\tau^a _0+\tau ^c_0}\phi=0 \label{tac0phi}
\eeq
so that 
\bea
\Pi _c^{\rm ex}&=& (-1)^{\tau ^c_0} \phi\left( 2[ \epsilon _1+ (-1)^{\tau ^c_2}\epsilon_4]
+[\tilde {\epsilon} _1+ (-1)^{\tau ^c_2}\tilde{\epsilon}_4]  \right)  \\
&=&{\Pi _c^{\rm ex}}' \label{cexaaa4}
\eea
The only other possibility is that $\tau ^c_1=1 \bmod 2$ and $|p_c|=2$. Then $2j+1=7$, and 
\beq
a \cap c, \ a \cap c' \geq 8
\eeq
which violate (\ref{akapineq}) and so are unacceptable.

Our conclusion is that to get the required quark-singlet spectrum, we must ensure that none of them arise at intersections with the $d$-type stack, and that they all arise from intersections with $c$ and $c'$. 
This requires that $d$ is given by  (\ref{Adp1111}), (\ref{adex}) and (\ref{aex4aaa}) with (\ref{tad2}) and  (\ref{tad0chi}).  It then follows that 
\beq
\#({\bf S}_d)=\frac{1}{2}(d \cap d'-d \cap \Pi _{\rm O6})=0
\eeq
 and there are no lepton singlets on $d$. Similarly, the $c$ stack is given by
\bea
2c&=& 2c'\\
&=&(2 +(-1)^{\tau ^c_1}) \rho _1+(-1)^{\tau ^c_0}\phi\left( 2[ \epsilon _1+ (-1)^{\tau ^c_2}\epsilon_4] 
+[\tilde {\epsilon} _1+ (-1)^{\tau ^c_2}\tilde{\epsilon}_4] \right)
\eea
with the constraint (\ref{tac0phi}). It follows that, 
\beq
\#({\bf S}_c)=0
\eeq
and there are no lepton singlets on $c$ either. Finally, 
 we find
\bea
d \cap c= d \cap c'&=&\frac{3}{2}[2+(-1)^{\tau ^c_1}+(-1)^{\tau ^d_0+\tau ^d_0}\chi\phi] \\
 &=&3 
\eea
using (\ref{tad0chi}) and (\ref{tac0phi}). 
It follows, using (\ref{yd1111}), that at these intersections we get  the three standard-model charged lepton singlets $3\ell^c_L$ plus three neutral lepton singlets  $3\nu ^c_L$ for either choice  of $y_c$. A similar analysis for the other 
solutions displayed in Table \ref{aaa1} yields the same conclusions, and the same physics. The first solution in the Table 
was the one used to illustrate our conclusions in the Erratum to \cite{Bailin:2007va}.

\subsection{Solution with ${\rm Im} \ U_3=-\sqrt{3}$} \label{ssaaa2}
Consider next the solution presented in Table \ref{aaa2}.
The supersymmetry constraint with this value of ${\rm Im} \ U_3$ on this lattice,
 together with the consistency condition $A^{\kappa} _1A^{\kappa} _6= A^{\kappa} _3A^{\kappa} _4$,  allows the same three classes of branes as were found for the lattice with ${\rm Im} \ U_3=-\frac{1}{\sqrt{3}}$, namely those characterised by equations (\ref{cAAA}), (\ref{dAAA}), ..., (\ref{enmAAA}). 

As noted in \S \ref{saaa2}, the $SU(2)_L$ stack $b$ is identical to that given in 
(\ref{pibex456}) for the three solutions in the bottom half of Table \ref{aaa1}.
 However, in this case, the supersymmetry constraints allow solutions of (\ref{bkbk1}) only when $\kappa$ is $d$-type with 
\bea
(A^d_1,A^d_3,A^d_4,A^d_6)&=&(1,1,1,1) \label{Adp1111II} \\
{\rm or} \ &=&(0,-1,0,1) \label{Adp0101}
\eea
These are just orientifold duals of each other, so that we need only consider the first possibility.
It  requires that
\beq
(n^d_1,m^d_1)= \eta \chi(-1,1), \ (n^d_2,m^d_2)=  \chi(1,-1), \ (n^d_3,m^d_3)= \eta (1,1) \label{nkdmkdII}
\eeq 
which gives the same intersection numbers as in (\ref{bdbd1n}), and we must  take $y_d$ as given in (\ref{yd1133}).
Again, we need at least one further $U(1)$ stack $\kappa$ satisfying (\ref{bkap11}). The solution is identical to that 
given in (\ref{Ac1000}) and (\ref{n1m1c2}).

Next we  determine the quark-singlet states that arise on the $SU(3)_{\rm colour}$ stack $a$ , and at its intersections 
with $d$ and $c$. 
First,  since $ \#({\bf S}_a)=0$, it follows that
\beq
\#({\bf A}_a)= a \cap \Pi_{\rm O6}=2
\eeq
 Thus there are $2d^c_L$ quark-singlet states on $a$, and we require one further $d^c_L$ and $3u^c_L$ from the intersections of $a$ with the $U(1)$ stacks $d$ and $c$ . 
$d$  is specified  in (\ref{Adp1111II}) with (\ref{nkdmkdII}), then 
\beq
(a \cap d, a \cap d')=(0,1)+(-1)^{\tau ^a _0 +\tau ^d_0} \chi[1+(-1)^{\tau ^a _2+\tau ^d_2}]\left((-1)^{\tau ^d_1}-1 ,-(-1)^{\tau ^d_1}\right)
\eeq
With $y_d$ given by (\ref{yd1133}), the states at $a \cap d$ are ${\bf 3}_{Y=\frac{2}{3}}$, while those at $a \cap d'$ are ${\bf 3}_{Y=-\frac{1}{3}}$. Thus, to avoid vector-like quark-singlet matter, we require that $a \cap d,a\cap d' \leq 0$. The only acceptable solution is therefore when $a \cap d=0=a \cap d'$. 
Thus, the only possibility is to get all of the required quark-singlets from intersections with $c$. 
With $c$ given by  (\ref{Ac1000}) and (\ref{nc2mc22}), we find
\beq
(a \cap c, a \cap c')=-\frac{1}{2}(2j+1,2j+1)+\frac{1}{2}(-1)^{\tau ^a _0 +\tau ^c_0} \left(p_c[2-(-1)^{\tau ^c_1}]-\phi ,-p_c[2-(-1)^{\tau ^c_1}]-\phi \right)
\eeq
Hence
\beq
a \cap c- a \cap c'=(-1)^{\tau ^a _0 +\tau ^c_0}p_c[2-(-1)^{\tau ^c_1}] 
\eeq
and  to avoid vector-like matter
\beq
p_c=0 \ {\rm or} \ |p_c|=2
\eeq
the latter only being possible when $\tau ^c_1=0 \bmod 2$. However, $p_c=0$ gives $a \cap c = a \cap c'$ which cannot yield all of the missing quark singlets. The alternative requires that $2j+1=p_c^2+3=7$ so that
\beq
a \cap c+ a \cap c'=-7+(-1)^{\tau ^a _0 +\tau ^c_0}\phi \geq -6
\eeq
Thus, in this case we can{\em not} obtain the required quark-singlet spectrum.
\subsection{Solutions with ${\rm Im} \ U_3=-2/\sqrt{3}$ and ${\rm Im} \ U_3=-1/2\sqrt{3}$}
All three of the solutions given in Tables \ref{aaa3} and \ref{aaa4} violate the inequality (\ref{noAa}). They have $\#({\bf A}_a)= a \cap \Pi_{\rm O6}=-3$, corresponding to $3\bar{d}^c_L$ quark 
singlets on the $SU(3)_{\rm colour}$ stack $a$. Therefore, these 
models cannot  yield just the standard-model quark-singlet spectrum. 
The same objection applies to the solutions given in Table \ref{aba} for the {\bf ABA} lattice and Table \ref{bba} for the {\bf BBA} lattice.
\section{Solutions for {\bf BAA} lattice} \label{baa}
The treatment of the solutions found on the {\bf BAA} lattice proceeds very similarly to that in the previous section for the  solutions on the {\bf AAA} lattice. Both of the solutions given in Table \ref{baa3} for the case ${\rm Im} \ U_3=-2/\sqrt{3}$ and Table \ref{baa4}   for the case ${\rm Im} \ U_3=-\sqrt{3}/2$ violate (\ref{noAa}). They also have $\#({\bf A}_a)= a \cap \Pi_{\rm O6}=-3$, and so cannot  yield just the standard-model quark-singlet spectrum.
\subsection{Solution with ${\rm Im} \ U_3=-1/\sqrt{3}$} \label{ssbaa1}
This is given in Table \ref{baa1}, and its treatment is very similar to that given in \S \ref{ssaaa2}. In this case 
supersymmery constrains the bulk wrapping numbers to satisfy
\bea
(A^{\kappa}_1,A^{\kappa}_3,A^{\kappa}_4,A^{\kappa}_6) &=& (0,1,0,0) \bmod 2 \label{cBAA}\\
{\rm or} \ &=& (\theta,1,\theta,1) \bmod 2 \label{dBAA} \\
{\rm or} \ &=& (0,0,0,1) \bmod 2 \label{eBAA} 
\eea
where $\theta=0,1$. 
This restricts the allowed wrapping numbers $(n^{\kappa}_k,m^{\kappa}_k) $ to the cases
 \bea
 (n^{\kappa}_1,m^{\kappa}_1;n^{\kappa}_2,m^{\kappa}_2;n^{\kappa}_3,m^{\kappa}_3)&=& (1,0;1,1;1,0) \bmod 2  \label{cnmBAA}\\
 {\rm or} \ &=& (1,1;1,\theta;1,1) \bmod 2 \label{dnmBAA}  \\
{\rm or} \ &=& (0,1;0,1;0,1) \bmod 2 \label{enmBAA} 
\eea
We denote by $c$ stacks with wrapping numbers satisfying (\ref{cBAA}), (\ref{cnmBAA}),  by $d_{\theta}$ stacks with wrapping numbers satisfying (\ref{dBAA}), (\ref{dnmBAA}), and by $e$ stacks with wrapping numbers satisfying (\ref{eBAA}), (\ref{enmBAA}). 
With the $SU(2)_L$ stack $b$ given by (\ref{pib2}), to satisfy (\ref{bkbk1}) requires the existence of a $d_1$-type stack. 
The unique solution that avoids vector-like doublets again has the form given in (\ref{Adp1111II}) and (\ref{nkdmkdII})
which gives the same intersection numbers as in (\ref{bdbd1n}), and we must  take $y_d$ as given in (\ref{yd1133}).
Again, we need at least one further $U(1)$ stack $\kappa$ satisfying (\ref{bkap11}). 
However, on this occasion because the $SU(2)_L$ stack $b$  is 
 of type $c$, the extra $U(1)$ stack must be of type $e$ with
\beq
(A^e_1,A^e_3,A^e_4,A^e_6)=(2j+1)(0,0,0,1) \label{Ae0001}
\eeq
with the integer $j \geq 0$. This requires that
\beq
(n^e_3,m^e_3)=\eta(0,1)
\eeq
where $\eta = \pm 1$.
Then for every pair of (coprime) wrapping numbers of the form
\beq
(n^e_2,m^e_2)=\left([ (-1)^{\tau ^e_1}-1] p_e+[ (-1)^{\tau ^e_1}+1] \phi, p_e-(-1)^{\tau ^e_1}\phi]\right) \label{ne2me22} 
\eeq
with $p_e = 0 \bmod 2$, we are guaranteed to generate the required bulk wrapping numbers (\ref{Ae0001}) and intersection numbers
 provided that we choose $j$ such that
\beq
\left( \begin{array}{c}
             n^e_1 \\
             m^e_1
       \end{array} \right) = \frac{ \eta(2j+1)}{ [2-(-1)^{\tau ^e_1}]p_e^2+[2+(-1)^{\tau ^e_1}]}\left( \begin{array}{c}
                                              [1-(-1)^{\tau ^e_1}]p_e - [1+(-1)^{\tau ^e_1}]\phi \\
                                               p_e(-1)^{\tau ^e_1}+\phi
                                               \end{array} \right) \label{n1m1e2}
\eeq
are  also coprime integers.  

Since there are {\em two} antisymmetric representations ${\bf A}_a=\bar{\bf 3}_{Y=1/3}$ on the $SU(3)_{\rm colour}$ stack $a$, and these correspond to 
$2d^c_L$ quark singlets,   
 all of the remaining $d^c_L +3 u^c_L$ quark singlets must arise at intersections of $a$ with the $U(1)$ stacks $d$ and $e$. 
As in \S \ref{ssaaa2}, the only way to avoid vector-like quark singlets at the intersections of $a$ with $d$ and $d'$ is to 
ensure that $a \cap d =0=a \cap d'$. It follows from (\ref{Ae0001}) and (\ref{ne2me22}) that
\beq
a \cap e - a \cap e'=p_e(-1)^{\tau^a _0+\tau ^e_0}[2-(-1)^{\tau ^e_1}]
\eeq
and to get the missing quark singlets we therefore require that $|p_e|= 2$ and   $ \tau ^e_1=0 \bmod 2$. However, 
in that case, $j=3$ and 
\beq
a \cap e + a \cap e'=-7+(-1)^{\tau^a _0+\tau ^e_0}\phi
\eeq
where $\phi = \pm 1$. Hence, 
\beq
|a \cap e +a \cap e'| \geq 6
\eeq
so that we can{\em not} get just the standard-model quark-singlet spectrum.
\subsection{Solutions with ${\rm Im} \ U_3=-\sqrt{3}$} \label{ssbaa2}
Supersymmetry on this lattice again allows the same three classes as we found in \S \ref{ssbaa1}. The form of the $SU(2)_L$ 
stack $b$ for the first two solutions given in Table \ref{baa2} and (\ref{bex2baa}) again requires the existence of a $d$-type stack with 
\bea
(A^d_1,A^d_3,A^d_4,A^d_6)&=&(1,-1,-1,1) \label{Adp111121} \\
{\rm or} \ &=& (3,3,1,1) \label{Adp331121}
\eea
Proceeding as before these give 
\bea
(b \cap d , b \cap d')&=&(3,0), \ (0,-3), \ (2,-1) \ {\rm or} \ (1,-2)\\
{\rm or } \ &=& (-2,1) \  {\rm or} \ (-1,2)
\eea 
respectively. Hence, we must take 
\bea
y_d&=&\frac{1}{2}  \label{yd111121}\\
{\rm or } \ &=& -\frac{1}{2} \label{yd3311}
\eea
so that there are three (lepton) doublets $L$ having weak hypercharge $Y=-\frac{1}{2}$.  
On this occasion the fourth stack must be of type $c$ with 
\beq
(A^c_1,A^c_3,A^c_4,A^c_6) =(2j+1)(2,1,0,0) \label{Ac2100}
\eeq
Then 
\beq
(n^c_3,m^c_3)=\eta(1,0)
\eeq
where $\eta = \pm 1$, and  
for every pair of (coprime) wrapping numbers of the form
\beq
(n^c_2,m^c_2)=\left( \phi-(-1)^{\tau ^c_1} p_c , p_c+(-1)^{\tau ^c_1}\phi \right) \label{nc2mc22baa} 
\eeq
with $p_c=0 \bmod 2$, we are guaranteed to generate the required bulk wrapping numbers (\ref{Ac2100}) and intersection numbers
 provided that we choose $j$ such that
\beq
\left( \begin{array}{c}
             n^c_1 \\
             m^c_1
       \end{array} \right) = \frac{ \eta(2j+1)}{ [2-(-1)^{\tau ^c_1}]p_c^2+[2+(-1)^{\tau ^c_1}]}\left( \begin{array}{c}
                                              [2-(-1)^{\tau ^c_1}]p_c + [1+2(-1)^{\tau ^c_1}]\phi \\
                                               \phi[1-(-1)^{\tau ^c_1}]-p_c[1+(-1)^{\tau ^c_1}]
                                               \end{array} \right) \label{n1m1c2baa}
\eeq
are  also coprime integers. 

Consider the $SU(3)_{\rm colour}$ stack $a$ for the first solution presented in Table \ref {baa2}. Since there are { no} antisymmetric representations  on $a$, 
 all of the  $3d^c_L +3 u^c_L$ quark singlets must arise at intersections of $a$ with the $U(1)$ stacks $d$ and $c$. If we use the bulk part of $d$,  given in (\ref{Adp111121}), then the only acceptable result is  $a \cap d = 0 = a \cap d'$. 
If instead we use  (\ref{Adp331121}), then there is the additional possibility  
that $(a \cap d, a \cap d')=(-3,0)$, which corresponds to $3 u^c_L$ quark singlets at the intersections of $a$ with $d$. 
However, it follows from (\ref{nc2mc22baa}) that $a \cap c- a \cap c'=0 \bmod 2$, which means that  we can{\em not} get $3d^c_L$ quark-singlet states from the intersections of $a$ with $c$ and $c'$. Thus, as in \S \ref{ssaaa1}, 
we require that $a \cap d=0=a \cap d'$, and
we must get 
all of the quark-singlet states from the intersections with the fourth stack. The former requires that
\beq
(-1)^{\tau ^a_0+\tau ^d_0} \chi=\pm1 \label{ac0chi}
\eeq
where $n^d_2=\chi=\pm 1$,  the upper sign in (\ref{ac0chi}) corresponds to (\ref{Adp111121}), and the lower to (\ref{Adp331121}).   The latter requires that
\beq
p_c=0
\eeq
and hence that
\beq
j=0
\eeq
Then 
\beq
a \cap c=a \cap c'=-\frac{3}{2}[1+(-1)^{\tau ^a_0+\tau ^c_0+1}\phi]
\eeq
where $\phi=\pm 1$, so that $(a \cap c, a \cap c')=(-3,-3)$ 
provided that
\beq
(-1)^{\tau ^a_0+\tau ^c_0+1}\phi=1 \label{ad0phi}
\eeq
As before, no lepton-singlet states arise as symmetric representations ${\bf S}_d$ or ${\bf S}_c$ on the $U(1)$ stacks $d$ and $c$, and we find
\bea
d \cap c=d \cap c'&=&\pm\frac{3}{2}[1 \mp(-1)^{\tau ^d_0+\tau ^c_0+1}\chi\phi] \\
&=& \pm 3
\eea
again with the upper sign corresponding to (\ref{Adp111121}) and (\ref{yd111121}) and the lower to (\ref{Adp331121}) and (\ref{yd3311}); the last line follows using (\ref{ac0chi}) and (\ref{ad0phi}).
Either way we again  get three charged-lepton singlets $3\ell ^c_L$ plus three neutral-lepton singlets $3\nu^c_L$ for either choice of $y_c=\pm \frac{1}{2}$.

The treatment of the second solution in Table \ref{baa2} is almost the same. The two solutions have the same $SU(2)_L$ stack $b$, so that to get the correct $3L+\bar{H}+H$ doublet content, we need $U(1)$ stacks $d$ and $c$ of  the same general form as just found. The two solutions
 differ only in the form of the $SU(3)_{\rm colour}$ stack $a$. However, it turns out that
\beq
(\Pi _a^{\rm bulk}\cap \Pi _d^{\rm bulk},\Pi _a^{\rm bulk}\cap {\Pi _d^{\rm bulk}}')=
(\tilde{\Pi} _a^{\rm bulk}\cap {\Pi _d^{\rm bulk}}',\tilde{\Pi} _a^{\rm bulk}\cap {\Pi _d^{\rm bulk}})
\eeq
where $\tilde{\Pi} _a^{\rm bulk \ (ex)}$ is the bulk  (exceptional) part of $a$ in the first solution.
 Further, as noted after (\ref{piaex3baa}), 
\beq
\Pi _a^{\rm ex}=-\mathcal{R}{\tilde{\Pi} _a}^{\rm ex}
\eeq
It follows that
\beq
(\Pi _a^{\rm ex}\cap \Pi _d^{\rm ex},\Pi _a^{\rm ex}\cap {\Pi _d^{\rm ex}}')=
(\tilde{\Pi} _a^{\rm ex}\cap {\Pi _d^{\rm ex}}',\tilde{\Pi} _a^{\rm ex}\cap {\Pi _d^{\rm ex}})
\eeq
and hence that
\beq
(a \cap d, a \cap d')=(\tilde{a} \cap d', \tilde{a} \cap d)
\eeq
where $\tilde{a}$ denotes the full fractional $SU(3)$ stack in the first solution. Since the only acceptable solution 
 was $(\tilde{a} \cap d', \tilde{a} \cap d)=(0,0)$, we conclude that the same is the case for this solution and that $d$ 
 has exactly the same form as for the first solution. 
Likewise, 
\beq
\Pi _a^{\rm bulk}\cap \Pi _c^{\rm bulk}=\Pi _a^{\rm bulk}\cap {\Pi _c^{\rm bulk}}'=(2j+1)(-6) 
\eeq
exactly the same as in the first solution. Thus the  argument given above for $d$ follows again for $c$, and $c$ too  
has exactly the same form as before. 
Consequently  the physics of the two solutions is identical.

 A similar relationship exists between the two  solutions in the bottom half of Table \ref{baa2}. Both have the same $SU(2)_L$ stack $b$ as the solution in \S \ref{ssbaa1}. 
To get the correct $3L+\bar{H}+H$ doublet content, we therefore need $U(1)$ stacks $d$ and $e$ of  the same general form as found there. It turns out that the requirements of supersymmetry and the absence of vector-like doublets forces $d$ to have 
the same form (\ref{Adp111121}) or (\ref{Adp331121}) as found for the first and second solutions, and the same is true of the exceptional part. Likewise, the form of $e$ is as we found in \S \ref{ssbaa1} and  specified in (\ref{Ae0001}) and (\ref{ne2me22}). The only acceptable solutions again require that all quark-singlet states arise at intersections of the $SU(3)_{\rm colour}$ stack $a$ with $e$ and $e'$, which in turn requires that $p_e=0=j$. Then there are no symmetric representations on $d$ or $e$ and once again we get $3\ell ^c_L+3 \nu^c_L$ from the intersections of the $U(1)$ stacks. These two solutions therefore constitute a different realisation of the same physics as in the two models in the upper half of Table \ref{baa2}.

\section{Tadpole cancellation }
The cancellation of RR tadpoles, and hence of the NSNS tadpoles, requires that the overall homology class of the D6-branes and the O6-planes vanishes \cite{Blumenhagen:2000wh,Blumenhagen:2002wn,Blumenhagen:2002vp}:
\beq
\sum _{\kappa} N_{\kappa} (\kappa + \kappa')-4\Pi _{\rm O6}=0 \label{RR}
\eeq
where the sum is over all D6-brane stacks $\kappa$. Both bulk and exceptional parts are required to cancel separately. Since 
$\Pi _{\rm O6}$ has no exceptional part,  the contributions from the exceptional parts $\Pi _{\kappa} ^{\rm ex}$  of the various stacks $\kappa$ must cancel among themselves. In our case, $\kappa$ ranges over four stacks: the $SU(3)_{\rm colour}$ stack $a$, the $SU(2)_L$ stack $b$, and  the two $U(1)$ stacks that are always needed to ensure the correct, supersymmetric standard-model lepton/Higgs doublet content.

We only need consider the models in which $T^2_3$ is of {\bf A} type, since only (some of) these have been extended to give the standard-model spectrum. As previously noted, in all such models
 the $SU(2)_L$ stack $b$ has the property  (\ref{bb1ex}), and it follows that there is no contribution from this stack to the exceptional part of (\ref{RR}). Thus tadpole cancellation can only occur if the contributions from the exceptional parts of the other (three) stacks cancel. In the case of the standard-model solution derived in \S \ref{ssaaa1},  the $SU(3)_{\rm colour}$ stack $a$ is of type $d$, as defined in (\ref{dAAA}) and (\ref{dnmAAA}), and therefore uses the fixed points $(16)$ or $(45)$ in $T^2_3$; then $\Pi _a^{\rm ex}$ involves $\epsilon _{1,6}$ and $\tilde{\epsilon}_{1,6}$ or $\epsilon _{4,5}$ and $\tilde{\epsilon}_{4,5}$, as found in (\ref{piaex1}). The $U(1)$ stack denoted $d$ is also of $d$-type (hence the nomenclature), so that $\Pi _d^{\rm ex}$ involves the same exceptional cycles. However, the fourth stack $c$ is of $c$-type, defined in (\ref{cAAA}) and (\ref{cnmAAA}). It therefore uses the fixed points $(14)$ or $(56)$ in $T^2_3$, and $\Pi _c^{\rm ex}$ involves $\epsilon _{1,4}$ and $\tilde{\epsilon}_{1,4}$ or $\epsilon _{5,6}$ and $\tilde{\epsilon}_{5,6}$. Clearly, there is no possibility that the contributions to (\ref{RR})  from the exceptional parts of $a$, $d$, and $c$ cancel. In fact, for  models in which $T^2_3$ is of {\bf A} type,  such a cancellation requires either that  all three stacks are of the same general type $c$, $d$ or $e$, or that they are all of different types.  In all of our standard-model solutions $T^2_3$ {\em is}  of {\bf A} type, and none of them has the property necessary for cancellation. 
 
 In principle there are two ways to remedy the situation. One possibilty is to add further branes designed to remove the discrepancy. However, these will certainly generate additional gauge symmetry, and probably extra matter. It might be possible to arrange that the unwanted extra matter and gauge interactions are hidden from the observable sector containing the standard model that we have previously obtained. However, this  approach is contrary to our objective which was to obtain {\em just} the standard-model spectrum, and we have not explored it further. The alternative is to introduce background fluxes  \cite{Derendinger:2004jn, Kachru:2004jr, Grimm:2004ua, Villadoro:2005cu, DeWolfe:2005uu, Camara:2005dc}. These do not generate unwanted gauge groups or matter, and, in any case,  may be needed to stabilise moduli and to break supersymmetry.  
The 7-form gauge field $C_7$ to which the D6-branes are electrically coupled also couples to certain background  field strengths. The generalised  tadpole cancellation condition, in the presence of background fields, may be derived \cite{Camara:2005dc} from the contribution of the RR fields to the (massive) type IIA supergravity action, supplemented by the  coupling of the D6-branes to   $C_7$: 
 \beq
 S_{\rm IIA}=-\frac{1}{2\kappa _{10}^2}\int \frac{1}{2} F_2 \wedge ^*\!\!F_2+... + \mu _6\sum _{\kappa}N_{\kappa}\int_{\mathcal{M}_4 \times \kappa }C_7 \label{S2a}
 \eeq
 where
 \bea
 2\kappa _{10}^2&=&(2\pi)^7{\alpha '} ^4 \\
 \mu _6&=&(2\pi)^{-6}{\alpha '}^{-7/2}
 \eea
with $2\pi\alpha '$ the inverse string tension. The sum over $\kappa$ is here understood to include the orientifold image $\kappa'$ of $\kappa$, as well as the ${\rm O6}$-plane coupled with charge $-4\mu _6$. 
 \beq
 F_2=dC_1+mB_2+\bar{F}_2
 \eeq
 is the field strength of the 1-form gauge field $C_1$ dual to $C_7$
 \beq
 dC_7=F_8:=^*\!\!dC_1
 \eeq
  $B_2$ is the Kalb-Ramond 2-form with field strength
  \beq
  H_3=dB_2+\bar{H}_3
  \eeq
and $\bar{H}_3$   and $\bar{F}_2$ are respectively  the possibly non-zero NSNS and RR background field strengths with which we are concerned. 
  Then
  \beq
  ^*\!F_2=dC_7+m^*\!B_2+^*\!\!\bar{F}_2
  \eeq
  and
  \bea
  F_2\wedge ^*\!\!F_2&=&F_2 \wedge (dC_7 +...) \nonumber \\
  &=&d(F_2 \wedge C_7)+C_7\wedge (mH_3-m\bar{H}_3+d\bar{F}_2)+...
  \eea 
The D6-brane contribution to (\ref{S2a}) may be rewritten in terms of the Poincar\'e dual  3-form $\delta (\kappa)$ of $\kappa$
\beq
\int_{\mathcal{M}_4 \times \kappa }\!\!C_7=\int C_7 \wedge \delta(\kappa)
\eeq
so that requiring the absence of tadpole terms in a general  background field strength gives
\beq
\frac{1}{4\kappa _{10}^2}(m\bar{H}_3-d\bar{F}_2)+\mu _6\sum _{\kappa}N_{\kappa}\delta(\kappa)=0
\eeq
or in terms of the usual homology classes
\beq
\frac{1}{4\kappa _{10}^2}\Pi_{m\bar{H}_3-d\bar{F}_2}+\mu _6\left(\sum _{\kappa}N_{\kappa}(\kappa+\kappa ')-4\Pi _{\rm O6}\right)=0 \label{RRflux}
\eeq
where $\Pi_{m\bar{H}_3-d\bar{F}_2}$ is the 3-cycle of which $m\bar{H}_3-d\bar{F}_2$ is the Poincar\'e dual, and we have now displayed explicitly the contributions from $\kappa '$ and the ${\rm O6}$-plane. 
In the absence of the background fields  $\bar{H}_3$ and $d\bar{F}_2$, we retrieve the usual tadpole cancellation constraint (\ref{RR}), which is {\em not} satisfied by any of our models. In principle, then, background fields are {\em necessary} for a consistent theory. However, the fluxes of these fields through closed cycles are quantised. In general the flux  of the background field strength $\bar{F}_{p+2}$ associated with a D$p$-brane satisfies
\beq
\mu _p\int_{\Sigma_{p+2}}\bar{F}_{p+2}=2\pi m_p
\eeq
where
\beq
\mu _p=(2\pi)^{-p}{\alpha '}^{-(p+1)/2}
\eeq
is the  (electric) charge of the D$p$-brane, $\Sigma_{p+2}$ is a closed $(p+2)$-cycle, and $m_p \in \mathbb{Z}$. The mass $m$ is also a flux, quantised as $F_0=m=(4\pi ^2 \alpha ')^{-1/2}n_0$ with $n_0 \in \mathbb{Z}$, and the flux of  $\bar{H}_3$ is quantised like $\bar{F}_3$. The question then is whether we can solve the Diophantine equations that follow from the generalised tadpole cancellation condition (\ref{RRflux}).

As noted previously, tadpole cancellation must be satisfied separately by the bulk and exceptional brane contributions to (\ref{RRflux}). This requires that $\Pi_{m\bar{H}_3-d\bar{F}_2}$ is $\mathcal{R}$-invariant. Thus we need the Poincar\'e duals of the $\mathcal{R}$-invariant combinations of the untwisted bulk 3-cycles $\rho_p, \ p=1,3,4,6$, and of the $\theta ^3$-twisted sector cycles $\epsilon _j, \  \tilde{\epsilon}_j , \  j=1,4,5,6$. 
\subsection{Untwisted sector}
The Poincar\'e dual $\eta $ of a  closed $3$-cycle $\rho$ in a 6-dimensional manifold $\mathcal{M}$ is a 3-form  satisfying \cite{BottTu}
\beq
\int _{\rho} i^*\omega_3 = \int_{\mathcal{M}} \omega_3 \wedge \eta   \label{poinc}
\eeq
for an arbitrary { closed} $3$-form $\omega_3$. $^*i$ is the pullback mapping the  3-form $\omega _3$ on $\mathcal{M}$ to a 3-form on $\rho$. 
 We may choose
\bea
\sigma _0 \equiv dz_1 \wedge dz_2 \wedge dz_3 \\
\sigma _1 \equiv dz_1 \wedge dz_2 \wedge d\bar{z}_3 \\
\sigma _2 \equiv d\bar{z}_1 \wedge d\bar{z}_2 \wedge dz_3 \\
\sigma _3 \equiv d\bar{z}_1 \wedge d\bar{z}_2 \wedge d\bar{z}_3
\eea
as the 4 elements of the basis of $H^{3}(T^6/\mathbb{Z}_6')$, the space of invariant untwisted 3-forms. This space is dual to the space $H_{3}(T^6/\mathbb{Z}_6')$ of (untwisted) 3-cycles. Then the Poincar\'e duals $\eta _p, \ p=1,3,4,6$ of the 3-cycles $\rho _p$  are  
\bea
\eta_1&=& \frac{3}{4i{\rm Vol}(\mathcal{M})}(\bar{e}_1\bar{e}_3\bar{e}_5\sigma_0-\bar{e}_1\bar{e}_3{e}_5\sigma _1+{e}_1{e}_3\bar{e}_5\sigma _2-e_1e_3e_5\sigma _3) \\
\eta_3&=& \frac{3}{4i{\rm Vol}(\mathcal{M})}(-\alpha \bar{e}_1\bar{e}_3\bar{e}_5\sigma_0+\alpha \bar{e}_1\bar{e}_3{e}_5\sigma _1+\alpha ^2{e}_1{e}_3\bar{e}_5\sigma _2-\alpha ^2e_1e_3e_5\sigma _3) \\
\eta_4&=&\frac{3}{4i{\rm Vol}(\mathcal{M})}(\bar{e}_1\bar{e}_3\bar{e}_5\bar{\tau}_3\sigma_0-\bar{e}_1\bar{e}_3{e}_5\tau _3\sigma _1+{e}_1{e}_3\bar{e}_5\bar{\tau}_3\sigma _2-e_1e_3e_5\tau _3\sigma _3)\\
\eta_6&=& \frac{3}{4i{\rm Vol}(\mathcal{M})}(-\alpha \bar{e}_1\bar{e}_3\bar{e}_5\bar{\tau}_3\sigma_0+\alpha \bar{e}_1\bar{e}_3{e}_5\tau _3\sigma _1+\alpha ^2{e}_1{e}_3\bar{e}_5\bar{\tau}_3\sigma _2-\alpha ^2e_1e_3e_5\tau _3\sigma _3)
\eea
where $e_{2k-1}$ and $e_{2k}$ are complex numbers defining the basis 1-cycles of $T^2_k$, and
\beq
{\rm Vol}(\mathcal{M})=\frac{3}{4}|e_1|^2|e_3|^2|e_5|^2 {\rm Im} \tau _3
\eeq
since $T^2_1$ and $T^2_2$ are $SU(3)$ lattices; $\tau _3 \equiv e_6/e_5$ is the complex structure of $T^2_3$.

On the {\bf AAA} lattice,
\beq
e_1=\bar{e}_1\equiv R_1, \ e_3=\bar{e}_3\equiv R_3, \ e_5=\bar{e}_5\equiv R_5, \ {\rm Re} \ \tau_3=0
\eeq
and  the only model that gives the standard-model spectrum on the {\bf AAA} lattice has
\beq
{\rm Im} \ \tau _3= -\frac{1}{\sqrt{3}}
\eeq
Then the Poincar\'e duals of  the $\mathcal{R}$-invariant 3-cycles $\rho_1$ and $\rho _4+2 \rho _6$ are
\bea
\eta_1&=&\frac{i\sqrt{3}}{R_1R_3R_5}(\sigma_0-\sigma_1+\sigma _2-\sigma_3)\\
\eta_4+2\eta _6&=&\frac{i\sqrt{3}}{R_1R_3R_5}(\sigma_0+\sigma_1-\sigma _2-\sigma_3)
\eea
 both of which are, like $\bar{H}_3$,  {\em odd} under the action of $\mathcal{R}$. Using the flux quantisation conditions we infer that
\beq
\bar{H}_3=-\frac{\pi ^2 \alpha '}{9}[n_3(\eta _4+2\eta_6)+3n_6\eta_1]
\eeq
where  $n_{3,6}$ are the integers associated with the flux of $ \bar{H}_3$ through the 3-cycles $\rho _{3,6}$. 

Similarly, for the solutions on the {\bf BAA} lattice, for which 
\beq
{\rm Im} \ \tau _3= -{\sqrt{3}}
\eeq
we find that the 3-forms dual to the $\mathcal{R}$-invariant 3-cycles $\rho_6$ and $\rho _3+2 \rho _1$ are
\bea
\eta_6&=&\frac{i}{R_1R_3R_5}(\sigma_0+\sigma_1-\sigma _2-\sigma_3)\\
\eta_3+2\eta _1&=&\frac{i}{R_1R_3R_5}(\sigma_0-\sigma_1+\sigma _2-\sigma_3)
\eea
Again, both  are  {\em odd} under the action of $\mathcal{R}$, like $\bar{H}_3$.
In this case flux quantisation gives
\beq
\bar{H}_3=\frac{\pi ^2 \alpha '}{9}[3n_1\eta_6+n_4(2\eta_1+\eta _3)]
\eeq
with $n_{1,4} \in \mathbb{Z}$ the integers associated with the flux of $ \bar{H}_3$  through $\rho_{1,4}$. 
\subsection{$\theta ^3$-twisted sector}
At each of the 16  fixed points $(k,\ell)$ with $ k,\ell=1,4,5,6$ of $\mathbb{Z}_2$ in $T^2_1 \times T^2_3$ there is a collapsed 2-cycle that we have denoted by $f_{k,\ell}$. 
To compute the Poincar\'e duals of these we need to blow up the metric at each fixed point using the Eguchi-Hanson $EH_2$ metric \cite{Lutken}
\beq
ds^2=g_{i\bar{j}}dz^id\bar{z}^j
\eeq\
where $i,j=1,3$
\beq
g_{i\bar{j}}=A(u) \delta _{i\bar{j}}+B(u)( z_i-Z_i)( \bar{z}_j-\bar{Z}_j)
\eeq
with
\beq
u \equiv |z^1-Z^1|^2+ |z^3-Z^3|^2
\eeq
and the fixed point $(k,\ell)$ is at $(z^1,z^3)=(Z^1,Z^3) \in T^2_1 \times T^2_3$. The functions $A$ and $B$ are given by
\bea
A(u) &\equiv& u^{-1}(\lambda ^4+u^2)^{1/2} \\
B(u)&=&A'(u)
\eea
For $u \gg \lambda ^2$, $A(u) \sim 1$ and the metric is flat. However, for $u \ll \lambda ^2$,
\beq
g_{i\bar{j}}\simeq \frac{\lambda^2}{u} \left(\delta _{i\bar{j}}-\frac{ ( z_i-Z_i)( \bar{z}_j-\bar{Z}_j)}{u} \right)
\eeq
and we see that $ds^2$ is invariant under $(z^i-Z^i) \rightarrow \rho e^{i\chi}( z^i-Z^i)$. Thus $ds^2$ depends only on two parameters, rather than four. We can make this explicit by changing variables:
\bea 
z^1-Z^1:=\sqrt{u} \cos \left(\frac{\theta}{2}\right) e^{i(\chi+\phi)/2}  \label{z1}\\
z^3-Z^3:=\sqrt{u} \sin \left(\frac{\theta}{2}\right) e^{i(\chi-\phi)/2}  \label{z2}
\eea
Then
\beq
ds^2=\frac{\lambda ^2}{4}(d\theta ^2 +\sin ^2 \theta d\phi ^2) \label{S2}
\eeq
which is the metric on $S^2$ localised at $(k,\ell)$. We may therefore characterise the collapsed 2-cycle
\beq
f_{k,\ell}=\{(z^1,z^3)  \ | \ z^1=Z^1+ \sqrt{u} \cos \left(\frac{\theta}{2}\right) e^{i(\chi+\phi)/2}, z^3=Z^3+\sqrt{u} \sin \left(\frac{\theta}{2} \right)e^{i(\chi-\phi)/2}, 0 <\theta <\pi/2, \ 0<\phi <2 \pi \} \label{fkl}
\eeq
Note, we only need $0 <\theta <\pi/2$ for a closed 2-cycle  because of the $\mathbb{Z}_2$ symmetry. 

There is also an harmonic $(1,1)$-form localised at each fixed point $(k,\ell)$:
\beq
e_{k,\ell}=\omega _{i,\bar{j}}dz^i\wedge d\bar{z}^j
\eeq 
where
\beq
\omega_{i\bar{j}}=\alpha(u) \delta _{i\bar{j}}+\beta(u)( z_i-Z_i)( \bar{z}_j-\bar{Z}_j)
\eeq
with
\bea
\alpha(u) &\equiv& u^{-1}(\lambda ^4+u^2)^{-1/2}\lambda^4  \\
\beta(u)&=&\alpha '(u)
\eea
Note that $\alpha= {\rm O}(u^{-2})$ and $u\beta= {\rm O}(u^{-2})$ as $u \rightarrow \infty$, so the $(1,1)$-form  is localised at  the fixed point, the origin $u=0$. 
For $u \ll  \lambda ^2$, 
\beq
\omega _{i\bar{j}}= \frac{\lambda^2}{u}\left(\delta _{i\bar{j}}-\frac{ ( z_i-Z_i)( \bar{z}_j-\bar{Z}_j)}{u}\right)
\eeq
Thus on $f_{k,\ell}$ defined (\ref{fkl}) we get
\beq
e_{k,\ell}=\frac{\lambda^2 i}{2}\sin \theta  \ d\theta \wedge d\phi
\eeq
and hence
\beq
\int _{f_{k,\ell}}e_{m,n}= \pi i \lambda^2\delta _{k,m}\delta_{\ell,n}
\eeq
Similarly, the localisation of the $(1,1)$-forms gives
\beq
\int _{{T^4}} e_{k,\ell} \wedge e_{m,n}=\lambda ^4\pi ^2 \delta_{k,m}\delta _{\ell,n}
\eeq

The action of $\theta$  on the collapsed 2-cycles $f_{i,j}$  is
\bea
f_{1,j} &\rightarrow & f_{1,j} \\
f_{4,j} \rightarrow f_{5,j} &\rightarrow & f_{6,j} \rightarrow f_{4,j} \label{thetafij}
\eea
for each $j=1,4,5,6$.  It follows that 
\bea
\epsilon_j \equiv (f_{6,j}-f_{4,j}) \otimes \pi _3 +(f_{4,j}-f_{5,j}) \otimes \pi _4 \\
\tilde{\epsilon}_j \equiv (f_{4,j}-f_{5,j}) \otimes \pi _3 +(f_{5,j}-f_{6,j}) \otimes \pi _4
\eea
are invariant (exceptional) 3-cycles. In contrast, the localised harmonic $(1,1)$-forms $e_{i,j}$ transform as
\bea
e_{1,j} &\rightarrow & e_{1,j} \label{thetae1j}\\
e_{4,j} \rightarrow e_{6,j} &\rightarrow & e_{5,j} \rightarrow e_{4,j} \label{thetaeij}
\eea
and 
\bea
\omega _j & \equiv & [\alpha(e_{4,j}-e_{5,j})+(e_{5,j}-e_{6,j})]dz_2  \label{omj}\\
\tilde{\omega}_j&\equiv & [(e_{4,j}-e_{5,j})+\alpha(e_{5,j}-e_{6,j})]d\bar{z}_2 \label{tilomj}
\eea
may be taken respectively as the basis invariant twisted $(2,1)$- and $(1,2)$-forms in this sector. It follows that the Poincar\'e duals of $\epsilon _j$ and $\tilde{\epsilon}_j$ are respectively
\bea
\chi_j &=& \frac{1}{2\pi \lambda ^2 \alpha{\rm Vol}(T^2_2)}(-\alpha \bar{e}_3 \omega _j+e_3 \tilde{\omega}_j) \\
\tilde{\chi}_j &=& \frac{1}{2\pi \lambda ^2 \alpha{\rm Vol}(T^2_2)}( \bar{e}_3 \omega _j-\alpha e_3 \tilde{\omega}_j)
\eea
where
\beq
{\rm Vol}(T^2_2)=\frac{\sqrt{3}}{2}|e_3|^2
\eeq
On the {\bf AAA} lattice, the combinations $2\epsilon _j+\tilde{\epsilon}_j$ of exceptional 3-cycles are $\mathcal{R}$-invariant and their Poincar\'e duals 
$2\chi _j+\tilde{\chi}_j$ are  {\em odd}, like $\bar{H}_3$. The flux quantisation conditions then give
\beq
\bar{H}_3=-\frac{4\pi ^2 \alpha '}{3}\sum _j\hat{n}_j(2\chi _j+\tilde{\chi}_j)
\eeq
with $\hat{n}_j\in \mathbb{Z}$ the integer associated with the flux through $\epsilon _j$. On the {\bf BAA} lattice the corresponding result is
\beq
\bar{H}_3=\frac{4\pi ^2 \alpha '}{3}\sum _j\tilde{n}_j(\chi _j+2\tilde{\chi}_j)
\eeq
where  $\tilde{n}_j\in \mathbb{Z}$ is the integer associated with the flux through $\tilde{\epsilon} _j$. 

\vspace{1cm}

We may now use the foregoing results to rewrite the generalised tadpole cancellation condition (\ref{RRflux}). For the {\bf AAA} and {\bf BAA} lattices respectively, we get
\bea
-\frac{n_0}{36}[n_3(\rho_4 +2\rho_6)+3n_6\rho_1]+\frac{n_0}{3}\sum _j\hat{n}_j(2\epsilon _j+\tilde{\epsilon}_j)+\sum _{\kappa}N_{\kappa}(\kappa + \kappa')-4\pi _{\rm O6}=0 \\
\frac{n_0}{36}[{n}_4(\rho_3+2\rho_1)+3{n}_1\rho_6]-\frac{n_0}{3}\sum _j\tilde{n}_j(\epsilon _j+2\tilde{\epsilon}_j)+\sum _{\kappa}N_{\kappa}(\kappa + \kappa')-4\pi _{\rm O6}=0 
\eea
For the solution on the {\bf AAA} lattice given in \S \ref{ssaaa1} we find
\bea
\sum _{\kappa}N_{\kappa}(\kappa + \kappa ')-4\pi _{\rm O6}&=&[1+(-1)^{\tau ^c_1}] \rho _1+3(\rho _4+2\rho _6)+
(-1)^{\tau ^a_0 }\left(2[\epsilon _1+(-1)^{\tau ^a_2}\epsilon _6] 
+[\tilde{\epsilon}_1+(-1)^{\tau ^a_2}\tilde{\epsilon}_6]\right) \nonumber \\
&&+(-1)^{\tau ^c_0 }\left(2[\epsilon _1+(-1)^{\tau ^c_2}\epsilon _4] 
+[\tilde{\epsilon}_1+(-1)^{\tau ^c_2}\tilde{\epsilon}_4]\right)
\eea
Cancellation of the exceptional parts requires that $|n_0\hat{n}_4|=3$. Thus $|n_0|=1 \ {\rm or} \ 3$. In the first case, if $n_0=1$, then
\bea
&&\hat{n}_1= -3[(-1)^{\tau ^a_0 }+(-1)^{\tau ^c_0 }] \\
&&\hat{n}_4= -3(-1)^{\tau ^c_0 +\tau ^c_2 } \\
&&\hat{n}_6= -3(-1)^{\tau ^a_0 +\tau ^a_2 } \\
&&n_3=108 \\
&&n_6=12[1+(-1)^{\tau ^c_1}]
\eea
whereas, in the second case, if $n_0=3$, then
\bea
&& \hat{n}_1= -[(-1)^{\tau ^a_0 }+(-1)^{\tau ^c_0 }] \\
&&\hat{n}_4= -(-1)^{\tau ^c_0 +\tau ^c_2 } \\
&&\hat{n}_6= -(-1)^{\tau ^a_0 +\tau ^a_2 } \\
&&n_3=36 \\
&&n_6=4[1+(-1)^{\tau ^c_1}]
\eea

 Similarly, for the solution on the  {\bf BAA} lattice given in \S \ref{ssbaa2} we find 
\bea
\sum _{\kappa}N_{\kappa}(\kappa + \kappa ')-4\pi _{\rm O6}&=& 3\rho _6+
(-1)^{\tau ^a_0 }\left([\epsilon _1+(-1)^{\tau ^a_2}\epsilon _6] 
+2[\tilde{\epsilon}_1+(-1)^{\tau ^a_2}\tilde{\epsilon}_6]\right) \nonumber \\
&&+(-1)^{\tau ^c_0 }\left([\epsilon _1+(-1)^{\tau ^c_2}\epsilon _4] 
+2[\tilde{\epsilon}_1+(-1)^{\tau ^c_2}\tilde{\epsilon}_4]\right)
\eea
Thus to achieve the required cancellation of the exceptional parts, we must again take $|n_0|=1 \ {\rm or} \ 3$, and the  solutions are, if $n_0=1$ then
  \bea
&& \tilde{n}_1= 3[(-1)^{\tau ^a_0 }+(-1)^{\tau ^c_0 }] \\
&&\tilde{n}_4= 3(-1)^{\tau ^c_0 +\tau ^c_2 } \\
&&\tilde{n}_6= 3(-1)^{\tau ^a_0 +\tau ^a_2 } \\
&&n_1=36 \\
&&n_4=0
\eea
whereas if $n_0=3$, then
\bea
&& \tilde{n}_1= (-1)^{\tau ^a_0 }+(-1)^{\tau ^c_0 } \\
&&\tilde{n}_4= (-1)^{\tau ^c_0 +\tau ^c_2 } \\
&&\tilde{n}_6= (-1)^{\tau ^a_0 +\tau ^a_2 } \\
&&n_1=-12 \\
&&n_4=0
\eea
Thus, in both cases we {\em can} choose the background NSNS 3-form fieldstrength $\bar{H}_3$ so as to satisfy the tadpole cancellation conditions. There remains the possibilities that this could also be done using the background $d\bar{F}_2$ that derives from ``metric fluxes'', or by a combination of both. We have not  explored this further.
\section{Non-anomalous $U(1)$ groups}
Tadpole cancellation generally ensures  that  any anomalous $U(1)$ gauge symmetries are removed; the associated gauge boson acquires a string-scale mass via the generalised Green-Schwarz mechanism and the $U(1)$ survives only as a global symmetry of the theory. In any case, there remains the possibility that  non-anomalous $U(1)$s may survive as low-energy gauge symmetries, and indeed we require this to be the case for the $U(1)_Y$ associated with the weak hypercharge $Y$. 
The $U(1)$ gauge boson associated with a general linear combination 
of the $U(1)$ charges $Q_{\kappa}$
\beq
X = \sum _{\kappa} x_{\kappa} Q_{\kappa}
\eeq
whether anomalous or non-anomalous,
 does {\em not} acquire a mass via the Green-Schwarz mechanism provided that \cite{Honecker:2004kb,MarchesanoBuznego:2003hp,Ott:2003yv}
\beq
\sum_{\kappa}x_{\kappa}N_{\kappa}(\kappa - \kappa ')=0
\eeq

Consider again the model derived in \S \ref{ssaaa1}, deriving from the fourth entry in Table \ref{aaa1}. Using (\ref{aex4aaa}) we find  that 
\beq
2(a-a')=\rho_1+2\rho_3+3 \rho_4-3 (-1)^{\tau ^a_0}[ \tilde{\epsilon}_1+ (-1)^{\tau ^a_2} \tilde{\epsilon}_6]
\eeq 
Thus $a \neq a'$, which shows that the gauge boson of $U(1)_a$, associated with $Q_a$,  does not remain massless, and  $U(1)_a$ survives only as a global symmetry.  Since none of the quark-singlet states arise as antisymmetric representations on  the stack $a$, baryon number $B=\frac{1}{3}Q_a$. It follows that the global $U(1)_a$ symmetry is just baryon-number conservation.
Similarly, from (\ref{pibx1}) and (\ref{pibex456}) we find
\beq
2(b-b')=(-1)^{\tau ^b_0}[ \tilde{\epsilon}_1+ (-1)^{\tau ^b_2} \tilde{\epsilon}_5]
\eeq 
so that $b-b' \neq 0$, the gauge boson of $U(1)_b$ acquires a string-scale mass, and $U(1)_b$ also survives only as a global symmetry. From (\ref{Adp1111}) and (\ref{adex}) we find 
\beq
2(d-d')=-\rho_1-2\rho_3-3 \rho_4+3 (-1)^{\tau ^a_0}[ \tilde{\epsilon}_1+ (-1)^{\tau ^a_2} \tilde{\epsilon}_6]
\eeq 
Finally, from (\ref{Ac1000}) and (\ref{cexaaa4}) we find
\beq
c-c'=0
\eeq
For this solution, using (\ref{ya}), (\ref{yb}) and (\ref{yd1111}) 
\beq
Y= \frac{1}{6}Q_a+\frac{1}{2}Q_d \pm \frac{1}{2}Q_c
\eeq
It follows from these that $U(1)_Y$ {\em does} remain massless, as required. However, since $c=c'$, so too does $U(1)_c$. Thus, we have an unwanted $U(1)$ factor in the surviving gauge group, besides the  required $SU(3)_{\rm colour} \times SU(2)_L \times U(1)_Y$ of the standard model. 
It is easy to verify that $B-L$, where $B$ is baryon number and $L$ is lepton number, is given by
\beq
B-L=\frac{1}{3}Q_a+Q_d
\eeq
This gives the correct values for the quark and lepton states, and also ensures that the doublets that arise at the intersections of $b$ with $c$ and $c'$ have $B-L=0$. Thus these states {\em are} the (so far unobserved) Higgs doublets, and the unwanted $U(1)_{c}$ is just a linear combination of the massless $U(1)_Y$ with $U(1)_{B-L}$. 
The same defect is present in the other standard-model solutions, on both  the {\bf AAA} and {\bf BAA} lattices. 
\section{Conclusions}
The $\mathbb{Z}_6'$ orientifold is so far the only known compactification of Type IIA string theory that can accommodate intersecting supersymmetric stacks $a$ and $b$ (with $N_a=3$ and $N_b=2$) of (fractional) D6-branes  satisfying (\ref{abab1}), having no  matter in symmetric representations, and not too much in antisymmetric representations, on either stack. Stacks having these properties are a useful starting point if we are eventually to obtain {\em just} the spectrum of the supersymmetric Standard Model, although in principle $(a \cap b, a \cap b')=(0,3)$ or $(3,0)$ are also allowed. In a previous publication  \cite{Bailin:2006zf} we presented a number of examples possessing the former properties in cases in which $T^2_3$ is of {\bf B}-type, and in this paper we have obtained solutions when $T^2_3$ is of {\bf A}-type. We have also studied whether any of our solutions {\em can} be extended to give just the (supersymmetric)  standard-model spectrum by the addition of extra $U(1)$ stacks $c,d,...$ with $N_{c,d,...}=1$. In  all of the former cases, as detailed in \S \ref{nogot2b}, the answer is negative. 
The same is immediately true for the solutions found on the {\bf ABA} and {\bf BBA} lattices, since they have $\bar{d}^c_L$ quark-singlet states arising as antisymmetric matter on the $SU(3)_{\rm colour}$ stack $a$. 
However, we {\em have} found models that give the standard-model spectrum, always accompanied by three neutrino-singlet states $3\nu ^c_L$, on the {\bf AAA} and {\bf BAA} lattices. In all cases we require two $U(1)$ stacks to get the correct 
lepton/Higgs doublet content. Baryon number conservation survives as a global symmetry in all of our solutions. 
Also in all cases, though, besides the standard-model $SU(3)_{\rm colour} \times SU(2)_L \times U(1)_Y$  gauge group,  there is unavoidably an additional (non-anomalous) $U(1)$ factor surviving as a local, rather than a global, symmetry.  This is effectively  $U(1)_{B-L}$, the $U(1)$ associated with baryon number $B$ minus lepton number $L$. Assuming that  it can be broken, such a $U(1)$ is in principle useful \cite{Mohapatra:1980qe, Marshak:1979fm, Wetterich:1981bx, Khalil:2006yi, Khalil:2007dr, Khalil:2008kp} in linking the neutrino masses and non-baryonic dark matter.

 In the first instance, the solutions obtained when  $T^2_2$ is of {\bf A}-type do not satisfy the tadpole-cancellation conditions (\ref{RR}), so that they are not consistent configurations of D6-branes. However, the 7-form gauge potential $C_7$ associated with D6-branes also couples to the background NSNS 3-form field strength $\bar{H}_3$, and this leads to modification of the tadpole cancellation conditions in the presence of such flux. We have shown that it is possible to choose the background so that the modified tadpole cancellation conditions {\em are} satisfied. Nevertheless,  there remain unstabilised K\"ahler and dilaton moduli. It is known that  in principle these may be stabilised using RR, NSNS and metric fluxes \cite{Derendinger:2004jn, Kachru:2004jr, Grimm:2004ua, Villadoro:2005cu, DeWolfe:2005uu}.  Models 
 similar to the ones we have been discussing  can be 
 uplifted into ones with stabilised K\"ahler moduli using a ``rigid corset'' \cite{Camara:2005dc, Aldazabal:2006up}, which can be added to any RR tadpole-free assembly of D6-branes in order to stabilise all moduli. 
 Fluxes may also be necessary  to break supersymmetry. 
 So far, we have  only explored models in which both $T^2_1$ and $T^2_2$ are $SU(3)$ root lattices. Either or both could be $G_2$ root lattices, and the results presented here illustrate amply how different lattices give different physics. 
 We shall explore all of these possibilities in future work.
\section*{Acknowledgements} 
We are grateful to Mirjam Cveti\v{c}, Gabriele Honecker, Eran Palti and Fernando Quevedo for helpful discussions. This work was funded in part by PPARC.

\appendix 
\section{ Calculations of $(i^a_1,i^a_2)(j^a_1,j^a_2)\cap (i^b_1,i^b_2)(j^b_1,j^b_2)'$ on the {\bf AAA} lattice} \label{AAA}
\subsection{$(n^{a,b}_1, m^{a,b}_1)=(n^{a,b}_3, m^{a,b}_3)=(1,0) \bmod 2$}
\label{101}
\bea
(56)(14) &\cap &  (56)(14)' =(56)(56) \cap (56)(56)' =\nonumber \\
&=& (-1)^{\tau ^a _0 +\tau ^b _0 }2[1+(-1)^{\tau ^a_2 +\tau ^b_2}]\left[ (n^a_2n^b_2-m^a_2m^b_2) 
-(-1)^{\tau ^a _1 +\tau ^b_1}(n^a_2n^b_2+n^a_2m^b_2+m^a_2n^b_2) \right.+ \nonumber \\
&+& \left.[(-1)^{\tau ^a_1}+(-1)^{\tau ^b_1}](m^a_2m^b_2+n^a_2m^b_2+m^a_2n^b_2) \right] \\
(56)(14) &\cap &  (56)(56)' =(56)(56) \cap   (56)(14)' =0 
\eea
\subsection{$(n^{a,b}_1, m^{a,b}_1)=(n^{a,b}_3, m^{a,b}_3)=(1,1) \bmod 2$}
\label{111}
\bea
(45)(16) &\cap &  (45)(16)' = (45)(45) \cap (45)(45)' =\nonumber \\
&=& (-1)^{\tau ^a _0 +\tau ^b _0 }2[1+(-1)^{\tau ^a_2 +\tau ^b_2}]\left[m^a_2m^b_2+n^a_2m^b_2+m^a_2n^b_2 \right.+ \nonumber \\
&+&\left. [(-1)^{\tau ^a_1+1}+(-1)^{\tau ^b_1+1}](n^a_2n^b_2+n^a_2m^b_2+m^a_2n^b_2)+
 (-1)^{\tau ^a_1+\tau ^b_1}(n^a_2n^b_2-m^a_2m^b_2) \right] \\
(45)(16) &\cap &  (45)(45)' =0 
=(45)(45) \cap (45) (16)'
\eea
\subsection{$(n^{a,b}_1,m^{a,b}_1)=(n^{a,b}_3,m^{a,b}_3)=(0,1)  \bmod 2$} \label{011}
\bea
(46)(15) &\cap &  (46)(15)' =(46)(46) \cap (46)(46)' =\nonumber \\
&=& (-1)^{\tau ^a _0 +\tau ^b _0 }2[1+(-1)^{\tau ^a_2 +\tau ^b_2}]\left[m^a_2m^b_2+n^a_2m^b_2+m^a_2n^b_2 \right.- \nonumber \\
&-& \left. (-1)^{\tau ^a_1 +\tau ^b_1}(n^a_2n^b_2+n^a_2m^b_2+m^a_2n^b_2) 
+  [(-1)^{\tau ^a _1} +(-1)^{\tau ^b_1}](n^a_2n^b_2-m^a_2m^b_2)  \right] \\
(46)(15) &\cap &  (46)(46)' =0 =(46)(46) \cap (46)(15)'
\eea
\subsection{$(n^{a}_1, m^{a}_1)=(n^{a}_3, m^{a}_3)=(1,0) \bmod 2$, \ $(n^{b}_1, m^{b}_1)=(n^{b}_3, m^{b}_3)=(1,1) \bmod 2$} \label{10111}
$(n^a_2,m^a_2)(n^b_2,m^b_2)=(1,0)(0,1) \bmod 2$ and $(1,1)(1,1) \bmod2$. In the other 2 cases $f_{AB'}=2 \bmod 4$.
\bea
(56)(14) &\cap &  (45)(16)' =(-1)^{\tau ^a_2}(56)(14)\cap (45)(45)'= \nonumber \\
&=&(-1)^{\tau ^b_2} (56)(56) \cap (45)(45)'=(-1)^{\tau ^a_2+\tau ^b_2}(56)(56) \cap (45)(16)' =\nonumber \\
&=& (-1)^{\tau ^a _0 +\tau ^b _0 }2 \left[-(n^a_2n^b_2+n^a_2m^b_2+m^a_2n^b_2) 
+(-1)^{\tau ^a_1+\tau ^b_1}(m^a_2m^b_2+n^a_2m^b_2+m^a_2n^b_2) \right. + \nonumber \\
&+&\left. [(-1)^{\tau ^a_1}+(-1)^{\tau ^b_1}](n^a_2n^b_2-m^a_2m^b_2)           \right]  
\eea
\subsection{$(n^{a}_1, m^{a}_1)=(n^{a}_3, m^{a}_3)=(1,1) \bmod 2$, \ $(n^{b}_1, m^{b}_1)=(n^{b}_3, m^{b}_3)=(0,1) \bmod 2$} 
\label{11011}
In the 2 cases in which $(n^a_2,m^a_2)=(n^b_2,m^b_2) \bmod 2$ we find that $f_{AB'} = 2 \bmod 4$. In the other 2 cases it is 
$0 \bmod 4$.
\bea
(45)(16) &\cap &  (46)(15)' =(-1)^{\tau ^a_2+ \tau ^b_2}(45)(16)\cap (46)(46)'= \nonumber \\
&=& (45)(45) \cap (46)(46)'=(-1)^{\tau ^a_2+ \tau ^b_2}(45)(45) \cap (46)(15)' =\nonumber \\
&=& (-1)^{\tau ^a _0 +\tau ^b _0 }2 \left[(m^a_2m^b_2+n^a_2m^b_2+m^a_2n^b_2)[1+(-1)^{\tau ^a_1+\tau ^b_1}] +\right. \nonumber \\
&-&\left.  (-1)^{\tau ^a_1}(n^a_2n^b_2+n^a_2m^b_2+m^a_2n^b_2)+(-1)^{\tau ^b_1}(n^a_2n^b_2-m^a_2m^b_2)      \right]  
\eea
\subsection{$(n^{a}_1, m^{a}_1)=(n^{a}_3, m^{a}_3)=(0,1) \bmod 2$, \ $(n^{b}_1, m^{b}_1)=(n^{b}_3, m^{b}_3)=(1,0) \bmod 2$} \label{01101}
In this case $f_{AB'}=0 \bmod 4$ in the case that $(n^a_2,m^a_2)(n^b_2,m^b_2)=(1,1)(1,0) \bmod 2$, 
and $2 \bmod 4$ in the 3 other cases.
\bea
(46)(15) &\cap &  (56)(14)' =(-1)^{\tau ^a_2}(46)(15)\cap (56)(56)'= \nonumber \\
&=& (-1)^{\tau ^a_2+ \tau ^b_2}(46)(46) \cap (56)(56)'=(-1)^{\tau ^b_2}(46)(46) \cap (56)(14)' =\nonumber \\
&=& (-1)^{\tau ^a _0 +\tau ^b _0 }2 \left[-(n^a_2n^b_2+n^a_2m^b_2+m^a_2n^b_2)[1+(-1)^{\tau ^a_1+\tau ^b_1}] \right. + \nonumber \\
&+&\left. (-1)^{\tau ^a_1}(m^a_2m^b_2+n^a_2m^b_2+m^a_2n^b_2)+(-1)^{\tau ^b_1}(n^a_2n^b_2-m^a_2m^b_2)           \right]  
\eea
\section{ Calculations of $(i^a_1,i^a_2)(j^a_1,j^a_2)\cap (i^b_1,i^b_2)(j^b_1,j^b_2)'$ on the {\bf BAA} lattice} \label{BAA}
On the {\bf ABA} lattice the action of $\mathcal{R}$ on the bulk 3-cycles $\rho _p \ (p=1,3,4,6)$ is the same as on the 
{\bf BAA} lattice. Consequently, the function $f_{AB'} \equiv \Pi _a^{\rm bulk} \cap {\Pi _b^{\rm bulk}}'$ that determines the bulk contribution to $a \cap b$
 is the same on the two lattices. In contrast, the action of 
 $\mathcal{R}$ on the exceptional 3-cycles $\epsilon _j, \tilde{\epsilon} _j, (j=1,4,5,6)$ differs by an overall
  sign on the two lattices. The relative sign of the bulk and exceptional contributions to  $a \cap b$ and $a \cap b'$ is 
  controlled by the overall phase $(-1)^{\tau ^a _0 +\tau ^b_0}$. Thus the calculations of these quantities on the 
  {\bf ABA} lattice may be obtained from those on the {\bf BAA} lattice by the replacement $\tau ^b_0 \rightarrow \tau ^b_0+1$, 
  but {\em only} in the expressions for $a \cap b'$. The results of calculations for the {\bf ABA} lattice may therefore be obtained immediately from those presented below. This does not mean that  we may obtain solutions for the fractional branes $a$ and $b$ having the required properties on the {\bf ABA} lattice  trivially from solutions on the {\bf BAA} lattice. 
The orientifold planes and the total homology class $\Pi _{\rm O6}$ are different 
  in the two cases, so that the constraint that there are no symmetric representations $\#(${\bf S}$_{a,b})=0$ is quite different. 
\subsection{$(n^{a,b}_1, m^{a,b}_1)=(n^{a,b}_3, m^{a,b}_3)=(1,0) \bmod 2$}
\bea
(56)(14) &\cap &  (56)(14)' =(56)(56) \cap (56)(56)' =\nonumber \\
&=& (-1)^{\tau ^a _0 +\tau ^b _0+1 }2[1+(-1)^{\tau ^a_2 +\tau ^b_2}]\left[ n^a_2n^b_2+n^a_2m^b_2+m^a_2n^b_2 
-(-1)^{\tau ^a _1 +\tau ^b_1}(m^a_2m^b_2+n^a_2m^b_2+m^a_2n^b_2) \right.+ \nonumber \\
&+& \left.[(-1)^{\tau ^a_1}+(-1)^{\tau ^b_1}](m^a_2m^b_2-n^a_2n^b_2) \right] \\
(56)(14) &\cap &  (56)(56)' =(56)(56) \cap   (56)(14)' =0 
\eea
\subsection{$(n^{a,b}_1, m^{a,b}_1)=(n^{a,b}_3, m^{a,b}_3)=(1,1) \bmod 2$}
\label{aba11}
\bea
(45)(16) &\cap &  (45)(16)' = (45)(45) \cap (45)(45)' =\nonumber \\
&=& (-1)^{\tau ^a _0 +\tau ^b _0 +1}2[1+(-1)^{\tau ^a_2 +\tau ^b_2}]\left[m^a_2m^b_2-n^a_2n^b_2  
+ (-1)^{\tau ^a_1+\tau ^b_1}(n^a_2n^b_2+n^a_2m^b_2+m^a_2n^b_2)\right.+ \nonumber \\
&+&\left. [(-1)^{\tau ^a_1+1}+(-1)^{\tau ^b_1+1}](m^a_2m^b_2+n^a_2m^b_2+m^a_2n^b_2)+
  \right] \\
(45)(16) &\cap &  (45)(45)' =0 
=(45)(45) \cap (45) (16)'
\eea
\subsection{$(n^{a,b}_1,m^{a,b}_1)=(n^{a,b}_3,m^{a,b}_3)=(0,1)  \bmod 2$} \label{aba01}
\bea
(46)(15) &\cap &  (46)(15)' =(46)(46) \cap (46)(46)' =\nonumber \\
&=& (-1)^{\tau ^a _0 +\tau ^b _0+1 }2[1+(-1)^{\tau ^a_2 +\tau ^b_2}]\left[m^a_2m^b_2-n^a_2n^b_2-
(-1)^{\tau ^a_1 +\tau ^b_1}(m^a_2m^b_2+n^a_2m^b_2+m^a_2n^b_2) \right.+ \nonumber \\
&+& \left. [(-1)^{\tau ^a _1} +(-1)^{\tau ^b_1}](n^a_2n^b_2+n^a_2m^b_2+m^a_2n^b_2)  \right] \\
(46)(15) &\cap &  (46)(46)' =0 =(46)(46) \cap (46)(15)'
\eea
\subsection{$(n^{a}_1, m^{a}_1)=(n^{a}_3, m^{a}_3)=(1,0) \bmod 2$, \ $(n^{b}_1, m^{b}_1)=(n^{b}_3, m^{b}_3)=(1,1) \bmod 2$} \label{aba1011}
\bea
(56)(14) &\cap &  (45)(16)' =(-1)^{\tau ^a_2}(56)(14)\cap (45)(45)'= \nonumber \\
&=&(-1)^{\tau ^b_2} (56)(56) \cap (45)(45)'=(-1)^{\tau ^a_2+\tau ^b_2}(56)(56) \cap (45)(16)' =\nonumber \\
&=& (-1)^{\tau ^a _0 +\tau ^b _0 +1}2 \left[-(m^a_2m^b_2+n^a_2m^b_2+m^a_2n^b_2) 
+(-1)^{\tau ^a_1+\tau ^b_1}(m^a_2m^b_2-n^a_2n^b_2) \right. + \nonumber \\
&+&\left. [(-1)^{\tau ^a_1}+(-1)^{\tau ^b_1}](n^a_2n^b_2+n^a_2m^b_2+m^a_2n^b_2)           \right]  
\eea
\subsection{$(n^{a}_1, m^{a}_1)=(n^{a}_3, m^{a}_3)=(1,1) \bmod 2$, \ $(n^{b}_1, m^{b}_1)=(n^{b}_3, m^{b}_3)=(0,1) \bmod 2$} 
\label{aba1101}
\bea
(45)(16) &\cap &  (46)(15)' =(-1)^{\tau ^a_2+ \tau ^b_2}(45)(16)\cap (46)(46)'= \nonumber \\
&=& (45)(45) \cap (46)(46)'=(-1)^{\tau ^a_2+ \tau ^b_2}(45)(45) \cap (46)(15)' =\nonumber \\
&=& (-1)^{\tau ^a _0 +\tau ^b _0+1 }2 \left[(m^a_2m^b_2-n^a_2n^b_2)[1+(-1)^{\tau ^a_1+\tau ^b_1}] +\right. \nonumber \\
&-&\left.  (-1)^{\tau ^a_1}(m^a_2m^b_2+n^a_2m^b_2+m^a_2n^b_2)+(-1)^{\tau ^b_1}(n^a_2n^b_2+n^a_2m^b_2+m^a_2n^b_2)      \right]  
\eea
\subsection{$(n^{a}_1, m^{a}_1)=(n^{a}_3, m^{a}_3)=(0,1) \bmod 2$, \ $(n^{b}_1, m^{b}_1)=(n^{b}_3, m^{b}_3)=(1,0) \bmod 2$} 
\label{aba0110}
\bea
(46)(15) &\cap &  (56)(14)' =(-1)^{\tau ^a_2}(46)(15)\cap (56)(56)'= \nonumber \\
&=& (-1)^{\tau ^a_2+ \tau ^b_2}(46)(46) \cap (56)(56)'=(-1)^{\tau ^b_2}(46)(46) \cap (56)(14)' =\nonumber \\
&=& (-1)^{\tau ^a _0 +\tau ^b _0 +1}2 \left[-(m^a_2m^b_2+n^a_2m^b_2+m^a_2n^b_2)[1+(-1)^{\tau ^a_1+\tau ^b_1}] \right. + \nonumber \\
&+&\left. (-1)^{\tau ^a_1}(m^a_2m^b_2-n^a_2n^b_2)+(-1)^{\tau ^b_1}(n^a_2n^b_2+n^a_2m^b_2+m^a_2n^b_2)           \right]  
\eea
\section{ Calculations of $(i^a_1,i^a_2)(j^a_1,j^a_2)\cap (i^b_1,i^b_2)(j^b_1,j^b_2)'$ on the {\bf BBA} lattice} \label{BBA}
\subsection{$(n^{a,b}_1, m^{a,b}_1)=(n^{a,b}_3, m^{a,b}_3)=(1,0) \bmod 2$}
\label{bba10}
\bea
(56)(14) &\cap &  (56)(14)' =(56)(56) \cap (56)(56)' =\nonumber \\
&=& (-1)^{\tau ^a _0 +\tau ^b _0 }2[1+(-1)^{\tau ^a_2 +\tau ^b_2}]\left[-( m^a_2m^b_2+n^a_2m^b_2+m^a_2n^b_2) 
+(-1)^{\tau ^a _1 +\tau ^b_1}(m^a_2m^b_2-n^a_2n^b_2) \right.+ \nonumber \\
&+& \left.[(-1)^{\tau ^a_1}+(-1)^{\tau ^b_1}](n^a_2n^b_2+n^a_2m^b_2+m^a_2n^b_2) \right] \\
(56)(14) &\cap &  (56)(56)' =(56)(56) \cap   (56)(14)' =0 
\eea

\subsection{$(n^{a,b}_1, m^{a,b}_1)=(n^{a,b}_3, m^{a,b}_3)=(1,1) \bmod 2$}
\label{bba11}
\bea
(45)(16) &\cap &  (45)(16)' = (45)(45) \cap (45)(45)' =\nonumber \\
&=& (-1)^{\tau ^a _0 +\tau ^b _0 }2[1+(-1)^{\tau ^a_2 +\tau ^b_2}]\left[n^a_2n^b_2+n^a_2m^b_2+m^a_2n^b_2  - \right. \nonumber \\
&-&\left. (-1)^{\tau ^a_1+\tau ^b_1}(m^a_2m^b_2+n^a_2m^b_2+m^a_2n^b_2)
+ [(-1)^{\tau ^a_1}+(-1)^{\tau ^b_1}](m^a_2m^b_2-n^a_2n^b_2)
  \right] \\
(45)(16) &\cap &  (45)(45)' =0 
=(45)(45) \cap (45) (16)'
\eea
\subsection{$(n^{a,b}_1,m^{a,b}_1)=(n^{a,b}_3,m^{a,b}_3)=(0,1)  \bmod 2$} \label{bba01}
\bea
(46)(15) &\cap &  (46)(15)' =(46)(46) \cap (46)(46)' =\nonumber \\
&=& (-1)^{\tau ^a _0 +\tau ^b _0 }2[1+(-1)^{\tau ^a_2 +\tau ^b_2}]\left[n^a_2n^b_2+n^a_2m^b_2+m^a_2n^b_2 \right. \nonumber \\
&+& \left. (-1)^{\tau ^a_1 +\tau ^b_1}(m^a_2m^b_2-n^a_2n^b_2)
- [(-1)^{\tau ^a _1} +(-1)^{\tau ^b_1}] (m^a_2m^b_2+n^a_2m^b_2+m^a_2n^b_2)  \right] \\
(46)(15) &\cap &  (46)(46)' =0 =(46)(46) \cap (46)(15)'
\eea
\subsection{$(n^{a}_1, m^{a}_1)=(n^{a}_3, m^{a}_3)=(1,0) \bmod 2$, \ $(n^{b}_1, m^{b}_1)=(n^{b}_3, m^{b}_3)=(1,1) \bmod 2$} \label{bba1011}

\bea
(56)(14) &\cap &  (45)(16)' =(-1)^{\tau ^a_2}(56)(14)\cap (45)(45)'= \nonumber \\
&=&(-1)^{\tau ^b_2} (56)(56) \cap (45)(45)'=(-1)^{\tau ^a_2+\tau ^b_2}(56)(56) \cap (45)(16)' =\nonumber \\
&=& (-1)^{\tau ^a _0 +\tau ^b _0 }2 \left[m^a_2m^b_2-n^a_2n^b_2 
+(-1)^{\tau ^a_1+\tau ^b_1}(n^a_2n^b_2+n^a_2m^b_2+m^a_2n^b_2) \right. + \nonumber \\
&+&\left. [(-1)^{\tau ^a_1+1}+(-1)^{\tau ^b_1+1}](m^a_2m^b_2+n^a_2m^b_2+m^a_2n^b_2)           \right]  
\eea
\subsection{$(n^{a}_1, m^{a}_1)=(n^{a}_3, m^{a}_3)=(1,1) \bmod 2$, \ $(n^{b}_1, m^{b}_1)=(n^{b}_3, m^{b}_3)=(0,1) \bmod 2$} 
\label{bba1101}
\bea
(45)(16) &\cap &  (46)(15)' =(-1)^{\tau ^a_2+ \tau ^b_2}(45)(16)\cap (46)(46)'= \nonumber \\
&=& (45)(45) \cap (46)(46)'=(-1)^{\tau ^a_2+ \tau ^b_2}(45)(45) \cap (46)(15)' =\nonumber \\
&=& (-1)^{\tau ^a _0 +\tau ^b _0 }2 \left[(n^a_2n^b_2+n^a_2m^b_2+m^a_2n^b_2)[1+(-1)^{\tau ^a_1+\tau ^b_1}] +\right. \nonumber \\
&+&\left.  (-1)^{\tau ^a_1}(m^a_2m^b_2-n^a_2n^b_2)-(-1)^{\tau ^b_1}(m^a_2m^b_2+n^a_2m^b_2+m^a_2n^b_2)      \right]  
\eea

\subsection{$(n^{a}_1, m^{a}_1)=(n^{a}_3, m^{a}_3)=(0,1) \bmod 2$, 
\ $(n^{b}_1, m^{b}_1)=(n^{b}_3, m^{b}_3)=(1,0) \bmod 2$} \label{bba0110}
\bea
(46)(15) &\cap &  (56)(14)' =(-1)^{\tau ^a_2}(46)(15)\cap (56)(56)'= \nonumber \\
&=& (-1)^{\tau ^a_2+ \tau ^b_2}(46)(46) \cap (56)(56)'=(-1)^{\tau ^b_2}(46)(46) \cap (56)(14)' =\nonumber \\
&=& (-1)^{\tau ^a _0 +\tau ^b _0 }2 \left[(m^a_2m^b_2-n^a_2n^b_2)[1+(-1)^{\tau ^a_1+\tau ^b_1}] \right. + \nonumber \\
&+&\left. (-1)^{\tau ^a_1}(n^a_2n^b_2+n^a_2m^b_2+m^a_2n^b_2)-(-1)^{\tau ^b_1}(m^a_2m^b_2+n^a_2m^b_2+m^a_2n^b_2)           \right]  
\eea

\end{document}